\documentclass[seceqn]{elsart}
\usepackage{amssymb,graphics}

\newcommand{\D}{\rlap{\hspace{0.2em}/}D}
\newcommand{\V}{{\makebox[0pt]{\hspace{0.4em}$\scriptstyle/$}v}}
\newcommand{\Tr}{\mathop{\mathrm{Tr}}\nolimits}
\newcommand{\tfrac}[2]{{\textstyle\frac{#1}{#2}}}
\begin{document}

\begin{frontmatter}
\begin{flushright}
\Large TTP02-11\\
\Large SFB/CPP-03-03
\end{flushright}
\title{Asymptotics of the perturbative series\\
for $f_{B^*}/f_B$}

\author{F.~Campanario}\ead{francam@particle.uni-karlsruhe.de},
\author{A.G.~Grozin}\ead{grozin@particle.uni-karlsruhe.de},
\author{T.~Mannel}\ead{Thomas.Mannel@physik.uni-karlsruhe.de}
\address{Institut f\"ur Theoretische Teilchenphysik, Universit\"at Karlsruhe}

\begin{abstract}
\begin{sloppypar}
We investigate the structure of the leading IR renormalon singularity
in the QCD/HQET matching coefficients for heavy-light quark currents
beyond the large-$\beta_0$ limit.
From this result, we derive the large-order behaviour
of the perturbative series for these coefficients,
and for ratios of meson matrix elements, such as $f_{B^*}/f_B$.
\end{sloppypar}
\end{abstract}

\end{frontmatter}

\section{Introduction}
\label{Intro}

The understanding of the structure of the perturbative series has advanced
considerably over the recent years, see the review~\cite{Be:99}.
It became clear that perturbative series are at best asymptotic,
not even Borel-summable.
Based on an analysis of singularities in the Borel plane,
one can obtain the behaviour of the perturbative series for large $L$,
where $L$ is the order of perturbation theory.
The nearest singularity determines the leading asymptotic behaviour.

Most of the previous investigations use the large-$\beta_0$ limit,
whose relation to the real QCD is unclear.
At the first order in $1/\beta_0$, singularities in the Borel plane are simple poles.
At the higher orders, they become branching points.
However, there is an approach~\cite{Pa:78,BBK:97} based on the renormalization group,
which yields results with a real basis in QCD.
Singularities in the Borel plane are branching points,
whose powers are determined by the relevant anomalous dimensions,
but normalization factors cannot be calculated.

Effective field theories make use of the fact that a large scale is present,
and physical quantities can be expanded in inverse powers of this large scale.
In Heavy Quark Effective Theory (HQET, see the textbook~\cite{MW:00}),
this scale is the heavy quark mass $m$.
Renormalon singularities in HQET were investigated in~\cite{BB:94,BSUV:94}.
Unlike in QCD, the HQET heavy-quark self-energy has an UV renormalon
at positive $u$, namely $u=\frac{1}{2}$, which leads to an ambiguity
in the residual mass term.

A typical matrix element in the full theory, QCD,
is expanded in $1/m$:
\begin{equation}
{<}j{>} = C {<}\tilde{\jmath}{>}
+ \frac{1}{2m} \sum B_i {<}O_i{>}
+ \mathcal{O}\left(\frac{1}{m^2}\right)
\label{Exp0}
\end{equation}
(see~(\ref{Exp})), with short-distance matching coefficients $C$, $B_i$,\dots{}
and long-distance HQET matrix elements ${<}\tilde{\jmath}{>}$, ${<}O_i{>}$,\dots{}

The QCD matrix element ${<}j{>}$ contains no renormalon ambiguities,
if the operator $j$ has the lowest dimensionality in its channel.%
\footnote{Otherwise, there may be several ultraviolet renormalons
on the positive half-axis,
leading to ambiguities of the order $\Lambda_{\mathrm{QCD}}^n$
times lower-dimensional matrix elements.}
In HQET, we separate short- and long-distance contributions.
In schemes without strict separation of large and small momenta,
such as $\overline{\mathrm{MS}}$,
this procedure artificially introduces
infrared renormalon ambiguities in matching coefficients
and ultraviolet renormalon ambiguities in HQET matrix elements.
When calculating matching coefficients $C$,\dots,
we integrate over all loop momenta, including small ones.
Therefore, they contain,
in addition to the main short-distance contributions,
also contributions from large distances,
where the perturbation theory is ill-defined.
They produce infrared renormalon singularities,
factorially growing contributions to coefficients of the perturbative series,
which lead to ambiguities $\sim\left(\Lambda_{\mathrm{QCD}}/m\right)^n$
in the matching coefficients $C$,\dots{}
Similarly, HQET matrix elements of higher-dimensional operators ${<}O_i{>}$,\dots{}
contain, in addition to the main large-distance contributions,
also contributions from short distances,
which produce ultraviolet-renormalon singularities.
They lead to ambiguities of the order $\Lambda_{\mathrm{QCD}}^n$
times lower-dimensional matrix elements (e.g., ${<}\tilde{\jmath}{>}$).
These two kinds of renormalon ambiguities
have to cancel in physical full QCD
matrix elements ${<}j{>}$~(\ref{Exp0})~\cite{NS:95,LMS:95}.

Although this has been shown explicitly only in the large-$\beta_0$ limit,
it is assumed to hold beyond this approximation.
Based on this assumption, one may obtain additional information
on the structure of the infrared renormalon singularities of matching coefficients,
based on ultraviolet renormalons in higher-dimensional matrix elements,
which are controlled by the renormalization group.
This model-independent approach was applied to some simple HQET problems:
the heavy-quark pole mass~\cite{Be:95}
and the chromomagnetic-interaction coefficient~\cite{GN:97}.

In the present paper, we investigate heavy-light quark currents.
The asymptotic behaviour of the perturbative series
for the leading QCD/HQET matching coefficients
(due to the nearest infrared renormalon)
was studied in~\cite{NS:95,BG:95,Ne:95}%
\footnote{Note a typo in~(4.8) of~\cite{BG:95}: denominators of both terms
with $a$ should be $2\pi$, not $\pi$.}
in the large-$\beta_0$ limit.
Here we go beyond this approximation,
by using the renormalization-group based method.
The results on $1/m$ expansions of QCD heavy-light currents
are collected in Sect.~\ref{HL}.
We show that the asymptotic behaviour of the perturbative series
for the matching coefficients for all currents
follows from just four distinct cases,
two spin-0 currents and two spin-1 ones.
These cases are considered in Sects.~\ref{S0} and~\ref{S1} in detail.
Ratios of meson matrix elements, such as $f_{B^*}/f_B$,
are given by the ratios of the corresponding matching coefficients
at the leading order in $1/m$.
The asymptotics of the perturbative series for this ratio
is discussed in Sect.~\ref{Conc}.
The large two-loop correction in this ratio was observed in~\cite{BG:95};
here we present model-independent results for higher orders
which continue this trend.

\section{Heavy-light currents in QCD and HQET}
\label{HL}

Heavy Quark Effective Theory (HQET, see the textbook~\cite{MW:00})
has greatly advanced our understanding of many problems in heavy quark physics.
Its Lagrangian is~\cite{EH:90a,EH:90b,FGL:91}
\begin{eqnarray}
&&L = \bar{h}_v iv\cdot D h_v + \frac{1}{2m} \left[ O_k + C_m(\mu) O_m(\mu) \right]
+ \mathcal{O}(1/m^2)\,,
\label{Lagr}\\
&&O_k = - \bar{h}_v D_\bot^2 h_v\,,\quad
O_m = \frac{1}{2} \bar{h}_v G_{\alpha\beta} \sigma^{\alpha\beta} h_v\,,
\nonumber
\end{eqnarray}
where $h_v=\rlap/v h_v$ is the heavy-quark field, and $D_\bot=D-v(v\cdot D)$.
Due to reparametrization invariance~\cite{LM:92}, 
the kinetic-energy operator $O_k$ is not renormalized,
and its coefficient is unity to all orders in perturbation theory.

The chromomagnetic-interaction coefficient $C_m(\mu)$
can only be computed perturbatively, by matching the amplitudes
of an appropriate scattering process in QCD and HQET.
Solving the renormalization-group equation, we obtain
\begin{equation}
C_m(\mu) = \hat{C}_m
\left(\frac{\alpha_s(\mu)}{\alpha_s(\mu_0)}\right)^{-\frac{\gamma_{m0}}{2\beta_0}}
K_{-\gamma_m}(\alpha_s(\mu))\,,
\label{RGCm}
\end{equation}
where $\alpha_s(\mu)$ is the QCD coupling with $n_l$ light flavours
in the $\overline{\mathrm{MS}}$ scheme,
\begin{eqnarray*}
&&\beta(\alpha_s) = - \frac{1}{2} \frac{d\log\alpha_s}{d\log\mu}
= \beta_0 \frac{\alpha_s}{4\pi}
+ \beta_1 \left(\frac{\alpha_s}{4\pi}\right)^2 + \cdots ={}\\
&&\left( \frac{11}{3} C_A - \frac{4}{3} T_F n_l \right) \frac{\alpha_s}{4\pi}
+ \left( \frac{34}{3} C_A^2 - 4 C_F T_F n_l - \frac{20}{3} C_A T_F n_l \right)
\left(\frac{\alpha_s}{4\pi}\right)^2 + \cdots
\end{eqnarray*}
and for any anomalous dimension
$\gamma(\alpha_s)=\gamma_0\alpha_s/(4\pi)
+\gamma_1(\alpha_s/(4\pi))^2\allowbreak+\cdots$
we define
\begin{equation}
K_\gamma(\alpha_s) = \exp \int_0^{\alpha_s}
\left(\frac{\gamma(\alpha_s)}{2\beta(\alpha_s)}-\frac{\gamma_0}{2\beta_0}\right)
\frac{d\alpha_s}{\alpha_s}
= 1 + \frac{\gamma_0}{2\beta_0}
\left(\frac{\gamma_1}{\gamma_0}-\frac{\beta_1}{\beta_0}\right)
\frac{\alpha_s}{4\pi} + \cdots
\label{Kdef}
\end{equation}
The anomalous dimension of the chromomagnetic operator
$O_m$ is~\cite{EH:90b,FGL:91,ABN:97,CG:97}
\begin{equation}
\gamma_m = 2 C_A \frac{\alpha_s}{4\pi}
+ \frac{4}{9} C_A (17 C_A - 13 T_F n_l) \left(\frac{\alpha_s}{4\pi}\right)^2
+ \cdots
\label{gammam}
\end{equation}
The full one-loop correction to $C_m$ has been calculated  in~\cite{EH:90b},
and the two-loop correction in~\cite{CG:97}.
It is convenient to choose
\begin{equation}
\mu_0 = e^{-5/6} m\,.
\label{mu0}
\end{equation}
We have (see~\cite{GN:97})
\begin{eqnarray}
&&\hat{C}_m = 1 + c_{m1} \frac{\alpha_s(\mu_0)}{4\pi} + \cdots\,,
\label{cm1}\\
&&c_{m1} = 2 C_F + \frac{5}{2} C_A
- \left( 3 C_F + \frac{55}{6} C_A \right) \frac{C_A}{\beta_0}
+ \left( 11 C_F + 7 C_A \right) \frac{C_A^2}{\beta_0^2}\,.
\nonumber
\end{eqnarray}

Operators of full QCD are expanded in $1/m$;
coefficients of such expansions are HQET operators
with appropriate quantum numbers.
In the present paper we shall consider heavy-light quark currents
$j_0=\bar{q}_0\Gamma Q_0=Z_\Gamma'(\alpha_s'(\mu'))j(\mu')$,
where $\Gamma$ is an antisymmetrized product of $n$ Dirac $\gamma$ matrices
\begin{equation}
\Gamma = \gamma_\bot^{[\alpha_1}\cdots\gamma_\bot^{\alpha_n]}
\quad\mathrm{or}\quad
\gamma_\bot^{[\alpha_1}\cdots\gamma_\bot^{\alpha_{n-1}]} \rlap/v\,,
\label{Gamma}
\end{equation}
which commutes or anticommutes with $\rlap/v$:
\[
\rlap/v \Gamma = \sigma \Gamma \rlap/v\,,\quad
\sigma = \pm 1\,,
\]
and $\gamma_\bot^\alpha=\gamma^\alpha-\rlap/vv^\alpha$.
Here $\alpha_s'(\mu)$ is the QCD coupling with $n_f=n_l+1$ flavours.
At the leading order in $1/m$, we have only a single HQET current
$\tilde{\jmath}_0=\bar{q}_0\Gamma h_{v0}=\tilde{Z}(\alpha_s(\mu))\tilde{\jmath}(\mu)$,
while to subleading order in $1/m$ we get 
\begin{equation}
j(\mu') = C_\Gamma(\mu',\mu) \tilde{\jmath}(\mu)
       + \frac{1}{2m} \sum_i B^\Gamma_i(\mu',\mu) O_i(\mu)
       + \mathcal{O}(1/m^2)\,.
\label{Exp}
\end{equation}

The solution of the renormalization-group equation for $C_\Gamma(\mu',\mu)$ is
\begin{equation}
C_\Gamma(\mu',\mu) = \hat{C}_\Gamma
\left(\frac{\alpha_s'(\mu')}{\alpha_s'(\mu_0)}\right)^{\frac{\gamma_{\Gamma0}'}{2\beta_0'}}
\left(\frac{\alpha_s(\mu)}{\alpha_s(\mu_0)}\right)^{-\frac{\tilde{\gamma}_0}{2\beta_0}}
K_{\gamma_\Gamma'}'(\alpha_s'(\mu'))
K_{-\tilde{\gamma}}(\alpha_s(\mu))\,.
\label{RGsol}
\end{equation}
Here $K'$ involves the $n_f$-flavour $\beta$-function $\beta'$.
The anomalous dimensions of $\tilde{\jmath}$ and $j$ are
\begin{eqnarray}
&&\tilde{\gamma} = - 3 C_F \frac{\alpha_s}{4\pi}
\label{gammaj}\\
&&\quad{} + C_F \left[ \left(-\frac{8}{3}\pi^2+\frac{5}{2}\right) C_F
+ \left(\frac{2}{3}\pi^2-\frac{49}{6}\right) C_A
+ \frac{10}{3} T_F n_l \right]
\left(\frac{\alpha_s}{4\pi}\right)^2 + \cdots
\nonumber\\
&&\gamma_\Gamma' = - 2 (n-1) (n-3) C_F \frac{\alpha_s}{4\pi}
\nonumber\\
&&\qquad{} \times \left\{1 +
\left[ \tfrac{1}{2} (5(n-2)^2-19) C_F - \tfrac{1}{3} (3(n-2)^2-19) C_A \right]
\frac{\alpha_s}{4\pi} \right\}
\nonumber\\
&&\quad{} - \tfrac{1}{3} (n-1) (n-15) C_F \beta_0'
\left(\frac{\alpha_s}{4\pi}\right)^2 + \cdots
\label{gammaG}
\end{eqnarray}
where $n$ is the number of $\gamma$ matrices in $\Gamma$~(\ref{Gamma})
(of course, the anomalous dimension of the vector current
vanishes to all orders).
The anomalous dimension
$\tilde{\gamma}$~\cite{VS:87,PW:88,JM:91,BG:91}
of the HQET current $\tilde{\jmath}$ does not depend on $\Gamma$
(the three-loop term can be found using the method of~\cite{G:00}).
The anomalous dimension $\gamma_\Gamma'$ of the QCD current is known to
two~\cite{BG:95} and three loops~\cite{Gr:00}.
%(for the scalar current, even at four loops~\cite{Ch:97,VLR:97};
%for the vector current, it vanishes to all orders).
The full one-loop corrections to $C_\Gamma(m,m)$ were obtained in~\cite{EH:90a},
and two-loop ones in~\cite{BG:95,G:98}.
The renormalization-group invariants $\hat{C}_\Gamma$
are given by perturbative series in $\alpha_s(\mu_0)$:
\begin{eqnarray}
&&\hat{C}_\Gamma = 1 + \sum_{L=1}^\infty c^\Gamma_L
\left(\frac{\alpha_s(\mu_0)}{4\pi}\right)^L\,,
\label{series}\\
&&c^\Gamma_1 = C_F \Biggl\{ \tfrac{3}{2} (n-2)^2 - \eta (n-2) - \tfrac{13}{4}
\nonumber\\
&&\quad{} + \left[ \left(-\tfrac{4}{3}\pi^2+\tfrac{23}{4}\right) C_F
+ \left(\tfrac{1}{3}\pi^2+8\right) C_A \right] \frac{1}{\beta_0}
- \tfrac{3}{2} \left(11C_F+7C_A\right) \frac{C_A}{\beta_0^2}
\nonumber\\
&&\quad{} + \left[ \tfrac{5}{2} \left((n-2)^2-5\right) C_F
- \tfrac{1}{3} \left(3(n-2)^2-4\right) C_A \right]
(n-1) (n-3) \frac{1}{\beta_0'}
\nonumber\\
&&\quad{} + \left(11C_F+7C_A\right) (n-1) (n-3)
\frac{C_A}{\beta_0^{\prime2}} \Biggr\}\,,
\nonumber
\end{eqnarray}
where $\eta=-\sigma(-1)^n$.
It will be possible to find $c^\Gamma_2$ from the known two-loop results
for $C_\Gamma(m,m)$ when $\tilde{\gamma}_2$ will be known.
In this paper, our aim is to investigate the behaviour of the
coefficients $c^\Gamma_L$ at $L\gg1$.

There are various prescriptions for handling $\gamma_5$
in dimensional regularization.
Multiplying $\Gamma$ by the anticommuting $\gamma_5^{\mathrm{AC}}$
does not change $C_\Gamma$.
This is not true for the 't~Hooft--Veltman $\gamma_5^{\mathrm{HV}}$.
QCD currents with $\gamma_5^{\mathrm{AC}}$ and $\gamma_5^{\mathrm{HV}}$
are related by finite renormalization factors~\cite{Tr:79,La:93}:
\begin{equation}
\left(\bar{q}\Gamma\gamma_5^{\mathrm{AC}}Q\right)_{\mu'} =
K'_{\gamma'_{\Gamma\gamma_5^{\mathrm{AC}}}-\gamma'_{\Gamma\gamma_5^{\mathrm{HV}}}}
(\alpha_s'(\mu'))
\left(\bar{q}\Gamma\gamma_5^{\mathrm{HV}}Q\right)_{\mu'}\,,
\label{Larin}
\end{equation}
where the anomalous dimensions $\gamma'_{\Gamma\gamma_5^{\mathrm{AC}}}=\gamma'_{\Gamma}$
and $\gamma'_{\Gamma\gamma_5^{\mathrm{HV}}}$ differ starting from two loops.
In HQET, both currents have the same anomalous dimension $\tilde{\gamma}$,
and hence the corresponding renormalization factor is unity.
Therefore, $C_{\Gamma\gamma_5^{\mathrm{HV}}}(\mu',\mu)$ differs from $C_\Gamma(\mu',\mu)$
only by $K'_{\gamma'_\Gamma}(\alpha_s'(\mu'))$ in~(\ref{RGsol}),
and $\hat{C}_{\Gamma\gamma_5^{\mathrm{HV}}}=\hat{C}_\Gamma$.
For $\sigma^{\alpha\beta}$, multiplication by $\gamma_5^{\mathrm{HV}}$
is just a Lorentz rotation, and does not change the anomalous dimension.
Therefore, $\left(\bar{q}\sigma^{\alpha\beta}\gamma_5^{\mathrm{AC}}Q\right)_{\mu'}=
\left(\bar{q}\sigma^{\alpha\beta}\gamma_5^{\mathrm{HV}}Q\right)_{\mu'}$,
and $C_{\sigma_\bot}(\mu',\mu)=C_{\gamma_\bot\V}(\mu',\mu)$~\cite{BG:95},
where $\sigma_\bot^{\alpha\beta}=\frac{i}{2}[\gamma_\bot^\alpha,\gamma_\bot^\beta]$.

There are 8 different Dirac matrices $\Gamma$~(\ref{Gamma}) in
4-dimensional space.
For our investigation of $\hat{C}_\Gamma$,
we can restrict ourselves to
\begin{equation}
\Gamma=1\,,\quad
\rlap/v\,,\quad
\gamma_\bot^\alpha\,,\quad
\gamma_\bot^\alpha\rlap/v\,,
\label{Gamma4}
\end{equation}
because the other 4 matrices can be obtained from~(\ref{Gamma4})
by multiplying by $\gamma_5^{\mathrm{HV}}$.

The $1/m$ term in the expansion~(\ref{Exp}) was first investigated in~\cite{FG:90},
where the one-loop anomalous dimension matrix of dimension-4 operators $O_i$ was found.
The full one-loop corrections to $B_i$ for vector currents
(and axial currents with anticommuting $\gamma_5$)
were given in~\cite{GH:91,Ne:94}.
Some general properties of the matching coefficients $B_i$
and the anomalous dimension matrix of $O_i$
following from reparametrization invariance and equations of motion
were established in~\cite{Ne:94}, and the 
two-loop anomalous dimension matrix was calculated in~\cite{AN:98,BNP:01}.

The sum in~(\ref{Exp}) includes the bilocal terms~\cite{Lu:90}
originating from the insertion of the subleading terms of
the Lagrangian~(\ref{Lagr})
\begin{equation}
C_\Gamma \int dx\,i\,T\left\{\tilde{\jmath}(0),O_k(x)+C_m O_m(x)\right\}\,.
\label{Biloc}
\end{equation}
The states $|B{>}$, $|B^*{>}$ in HQET matrix elements will be understood
as the eigenstates of the leading HQET Hamiltonian with $m=\infty$;
they are $m$-independent.

The terms in the sum with the derivative acting on the heavy-quark field
can be obtained from reparametrization invariance~\cite{Ne:94}.
Let $\Gamma=\gamma^{[\alpha_1}\cdots\gamma^{\alpha_n]}$.
It can be decomposed into the parts commuting and anticommuting with $\rlap/v$:
\[
\Gamma = \Gamma_+ + \Gamma_-\,,\quad
\Gamma_\pm = \frac{1}{2} \left( \Gamma \pm \rlap/v\Gamma\rlap/v \right)\,.
\]
The matrix element of the renormalized QCD current $\bar{q}\Gamma Q$
from the heavy-quark state with momentum $mv$
to the light-quark state with momentum 0 is
\begin{equation}
\tfrac{1}{2} \left(C_{\Gamma_+}+C_{\Gamma_-}\right)
\bar{u}_q \Gamma u(mv)
+ \tfrac{1}{2} \left(C_{\Gamma_+}-C_{\Gamma_-}\right)
\bar{u}_q \rlap/v \Gamma u(mv)\,.
\label{mv}
\end{equation}
In this equation, we may substitute $v\to v+k/m$:
\begin{eqnarray*}
&&\tfrac{1}{2} \left(C_{\Gamma_+}+C_{\Gamma_-}\right)
\bar{u}_q \Gamma u(mv+k)\\
&&\quad{} + \tfrac{1}{2} \left(C_{\Gamma_+}-C_{\Gamma_-}\right)
\bar{u}_q \left(\rlap/v+\frac{\rlap/k}{m}\right) \Gamma u(mv+k)\,.
\end{eqnarray*}
Using $u(mv+k)=\left(1+\rlap/k/(2m)\right)u_v(k)$,
we obtain the leading term~(\ref{mv}) plus
\[
\frac{1}{4m} \left[ \left(C_{\Gamma_+}+C_{\Gamma_-}\right)
\bar{u}_q \Gamma \rlap/k u_v
+ \left(C_{\Gamma_+}-C_{\Gamma_-}\right)
\bar{u}_q \left( \rlap/v\Gamma\rlap/k + 2\rlap/k\Gamma \right) u_v
\right]\,.
\]
Therefore, the $\bar{q}Dh_v$ terms in the sum in~(\ref{Exp}) are
\begin{equation}
\tfrac{1}{2} \left(C_{\Gamma_+}+C_{\Gamma_-}\right)
\bar{q} \Gamma i\D h_v
+ \tfrac{1}{2}\left(C_{\Gamma_+}-C_{\Gamma_-}\right)
\bar{q} \left( \rlap/v \Gamma i\D + 2 i\D \Gamma \right) h_v\,.
\label{Repar}
\end{equation}

\begin{figure}[ht]
\begin{center}
\begin{picture}(42,19)
\put(21,9.5){\makebox(0,0){\includegraphics{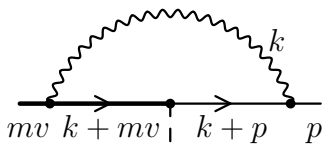}}}
\put(2,0){\makebox(0,0)[b]{$\strut{}mv$}}
\put(13,0){\makebox(0,0)[b]{$\strut{}k+mv$}}
\put(29,0){\makebox(0,0)[b]{$\strut{}k+p$}}
\put(40,0){\makebox(0,0)[b]{$\strut{}p$}}
\put(35,13){\makebox(0,0)[b]{$k$}}
\end{picture}
\end{center}
\caption{One-loop matching}
\label{Fig1}
\end{figure}

The coefficients of operators with the derivative acting on the
light-quark field are not determined by general considerations.
These coefficients appear first at the one-loop level.
We calculate the matrix element of the QCD current
from the heavy quark with momentum $mv$ to the light quark with momentum $p$
(with $p^2=0$),
expanded in $p/m$ to the linear term,
and equate it to the corresponding HQET matrix element.
In HQET, loop corrections contain no scale, and hence vanish
(except, possibly, massive-quark loops, which first appear
at the two-loop level).
The QCD matrix element is proportional to $\bar{u}(p)\Gamma(p,mv)u(mv)$,
where $\Gamma(p,mv)$ is the bare proper vertex function.
At one loop, it is given by Fig.~\ref{Fig1}.

If we assume nothing about properties of $\Gamma$,
then the term linear in $p$ has the structure
\begin{equation}
\bar{u}(p) \left[ \sum_i x_i L_i \Gamma R_i \right] u(mv)
\label{StrP}
\end{equation}
with $L_i\times R_i=p\cdot v\,1\times1$, $p\cdot v\,\rlap/v\gamma_\mu\times\gamma^\mu$,
$p\cdot v\,\gamma_\mu\gamma_\nu\times\gamma^\nu\gamma^\mu$,
$1\times\rlap/p$, $\rlap/v\gamma_\mu\times\gamma^\mu\rlap/p$.
The coefficients $x_i$ can be obtained, by solving a linear system,
from the double traces of Dirac matrices to the left from $\Gamma$ with $\bar{L}_j$
and those to the right from $\Gamma$ with $\bar{R}_j$,
with $\bar{L}_j\times\bar{R}_j=\rlap/v\rlap/p\times(1+\rlap/v)$,
$\gamma_\rho\rlap/p\times(1+\rlap/v)\gamma^\rho$,
$\gamma_\rho\gamma_\sigma\rlap/v\rlap/p\times(1+\rlap/v)\gamma^\sigma\gamma^\rho$,
$\rlap/v\rlap/p\times(1+\rlap/v)\rlap/p$,
$\gamma_\rho\rlap/p\times(1+\rlap/v)\rlap/p\gamma^\rho$.
Now we can take these double traces of the \emph{integrand} of Fig.~\ref{Fig1},
and express $x_i$ via scalar integrals.
Their numerators involve $(k\cdot p)^n$;
putting $k=(k\cdot v)v+k_\bot$ and averaging over $k_\bot$ directions
in the $(d-1)$-dimensional subspace, we can express them
via the factors in the denominator.

Now we assume
\begin{equation}
\rlap/v \Gamma = \sigma \Gamma \rlap/v\,,\quad
\sigma = \pm 1\,,\quad
\gamma_\mu \Gamma \gamma^\mu = 2 \sigma h \Gamma\,.
\label{GammaProp}
\end{equation}
For our $\Gamma$ matrices~(\ref{Gamma}),
\begin{equation}
h = \eta \left( n - \frac{d}{2} \right)\,,\quad
\eta = - \sigma (-1)^n\,.
\label{EtaH}
\end{equation}
Then~(\ref{StrP}) becomes
\begin{eqnarray}
&&\left[ x_1 + (x_2+2x_5)\cdot2h + x_3(2h)^2 \right]\,p\cdot v\, \bar{u}(p) \Gamma u(mv)
\nonumber\\
&&{} + \left[ x_4 - x_5\cdot2h \right] \bar{u}(p) \Gamma \rlap/p u(mv)\,.
\label{StrP2}
\end{eqnarray}
Performing the simple calculation, we obtain
\begin{eqnarray}
&&\bar{u}(p)\Gamma(p,mv)u(mv) =
\left[ 1 + C_F \frac{g_0^2 m^{-2\varepsilon}}{(4\pi)^{d/2}} \Gamma(\varepsilon)
c_\Gamma \right] \bar{u}(p) \Gamma u_v(0)
\nonumber\\
&&{} + \frac{1}{2m} C_F
\frac{g_0^2 m^{-2\varepsilon}}{(4\pi)^{d/2}} \Gamma(\varepsilon)
\left[ b^\Gamma_1\,p\cdot v\,\bar{u}(p) \Gamma u_v(0)
+ b^\Gamma_2 \bar{u}(p) \Gamma \rlap/p u_v(0) \right]\,,
\label{Vert1}
\end{eqnarray}
where
\begin{eqnarray*}
&&c_\Gamma = - \frac{(1-h)(d-2+2h)}{(d-2)(d-3)}\,,\\
&&b^\Gamma_1 = - 2 \frac{(d-2)(d-8)-(d-5)(d-4+2h)h}{(d-2)(d-3)(d-5)}\,,\\
&&b^\Gamma_2 = 2 \frac{d-2-h}{(d-2)(d-3)}\,.
\end{eqnarray*}
The zeroth-order coefficient $c_\Gamma$ has been found in~\cite{EH:90a,BG:95};
the first-order coefficients for components of the vector current
in~\cite{GH:91,Ne:94}.

For $\Gamma=1$, $\rlap/v$, the square bracket in~(\ref{Vert1})
becomes $b_\Gamma\,p\cdot v\,\bar{u}(p) \Gamma u_v(0)$,
with $b_1=b^1_1$, $b_\V=b^\V_1+2b^\V_2$.
We have checked these results
by taking the trace of the whole integrand of Fig.~\ref{Fig1}
with $(1+\rlap/v)\rlap/p$ and calculating the integrals.
For $\Gamma=\gamma_\bot^\alpha$, $\gamma_\bot^\alpha\rlap/v$,
the square bracket in~(\ref{Vert1}) becomes
$b_{\Gamma,1}\,p\cdot v\,\bar{u}(p)\gamma_\bot^\alpha u_v(0)
+b_{\Gamma,2}p_\bot^\alpha \bar{u}(p)u_v(0)$,
with $b_{\gamma_\bot,1}=b^{\gamma_\bot}_1$, $b_{\gamma_\bot,2}=2b^{\gamma_\bot}_2$,
$b_{\gamma_\bot\V,1}=b^{\gamma_\bot\V}_1+2b^{\gamma_\bot\V}_2$,
$b_{\gamma_\bot\V,2}=-2b^{\gamma_\bot\V}_2$.
We have checked these results by taking traces of the integrand with
$(1+\rlap/v)p_\alpha\rlap/p$ and $(1+\rlap/v)\gamma_\alpha\rlap/p$.
An additional strong check is provided by the Ward identity:
contracting the vertex function $\Gamma^\alpha(p,mv)$
(for $\Gamma=\gamma^\alpha$) with the momentum transfer $(mv-p)_\alpha$,
we obtain
\[
\Gamma^\alpha(p,mv) (mv-p)_\alpha = m \Gamma(p,mv) + \Sigma(mv)\,,
\]
where $\Gamma(p,mv)$ is the scalar vertex (for $\Gamma=1$),
and $\Sigma$ is the heavy-quark self-energy.
At the first order in $p$, this leads to
\begin{equation}
b_1 - b_\V = 2 \left( c_{\gamma_\bot} - c_\V \right)\,.
\label{Ward1}
\end{equation}
Our results~(\ref{Vert1}) satisfy this requirement.

In order to investigate the asymptotic behaviour
of the perturbative series~(\ref{series}),
we consider its Borel transform
\begin{equation}
S_\Gamma(u) = \sum_{L=1}^\infty \frac{c^\Gamma_L}{(L-1)!}\,
\left(\frac{u}{\beta_0}\right)^{L-1}\,,\quad
c^\Gamma_{n+1} = \left. \left(\beta_0\frac{d}{du}\right)^n
S_\Gamma(u) \right|_{u=0}\,.
\label{Borel}
\end{equation}
Formally one can invert this transformation and gets
\begin{equation}
\hat{C}_\Gamma = 1 + \frac{1}{\beta_0} \int_0^\infty S_\Gamma(u)\,
\exp\left(-\frac{4\pi}{\beta_0\alpha_s(\mu_0)}u\right)\,du\,.
\label{Laplace}
\end{equation}
However, if $S_\Gamma(u)$ has singularities on the integration contour
(which is the positive $u$ axis),
then the integral~(\ref{Laplace}) is not well-defined,
and the series~(\ref{series}) is not Borel-summable.
To deal with this singularities some prescription is needed, leading to
ambiguities in $\hat{C}_\Gamma$.

In the large-$\beta_0$ limit~\cite{BG:95},
\begin{eqnarray}
&&S_\Gamma(u) = C_F \Biggl\{ \frac{\Gamma(u)\Gamma(1-2u)}{\Gamma(3-u)}
\left[ 2 (n-2)^2 - 2 \eta (n-2) u + 3 u^2 + u - 5 \right]
\nonumber\\
&&\quad{} - \frac{2(n-2)^2-5}{2u} \Biggr\}
\label{Lb0}
\end{eqnarray}
(the results for the components of the vector current were obtained in~\cite{NS:95}).
Expanding this $S_\Gamma(u)$ in $u$ reproduces leading large-$\beta_0$ terms
in~(\ref{series}) (in particular, in $c^\Gamma_1$).
This Borel image has IR renormalon poles at $u>0$.
The pole nearest to the origin (and thus giving the largest renormalon ambiguity)
is situated at $u=\frac{1}{2}$:
\begin{equation}
S_\Gamma(u) = \frac{r_\Gamma}{\frac{1}{2}-u} +
(\mbox{regular terms at $u=\frac{1}{2}$})\,.
\label{SuLb0}
\end{equation}
This leads to an ambiguity in the sum~(\ref{Laplace}) of the series~(\ref{series}),
the natural measure of which is the residue at the pole:
\[
\Delta \hat{C}_\Gamma = \frac{r_\Gamma}{\beta_0} e^{5/6}
\frac{\Lambda_{\overline{\mathrm{MS}}}}{m}\,,
\]
where $\Lambda_{\overline{\mathrm{MS}}}$ is for $n_l$ flavours.
This is commensurate with the $1/m$ corrections in~(\ref{Exp}).
It is convenient to measure all such ambiguities in terms of
the UV renormalon ambiguity of $\bar{\Lambda}$~\cite{BB:94}
\begin{equation}
\Delta \bar{\Lambda} = - 2 C_F e^{5/6}
\frac{\Lambda_{\overline{\mathrm{MS}}}}{\beta_0}\,.
\label{DLbar}
\end{equation}
Then~\cite{BG:95},
\begin{equation}
\Delta \hat{C}_\Gamma = - \frac{1}{3}
\left[ 2 (n-2)^2 - \eta (n-2) - \frac{15}{4} \right]
\frac{\Delta\bar{\Lambda}}{m}\,.
\label{DCLb0}
\end{equation}

In the following Sections we shall see that beyond the large-$\beta_0$ limit
$S_\Gamma(u)$ has a branching point at $u=\frac{1}{2}$
rather than the simple pole~(\ref{SuLb0}).
We use the method based on renormalization group,
which will be explained in Sect.~\ref{S0}.

\section{Spin-0 currents}
\label{S0}

In the case of the currents with $\Gamma=1$, $\rlap/v$,
the leading term in the expansion~(\ref{Exp}) contains $\tilde{\jmath}=\bar{q}h_v$,
and the $1/m$ correction -- four operators
\begin{eqnarray}
&&O_1 = \bar{q} i\D h_v = i\partial_\alpha \left(\bar{q} \gamma^\alpha h_v\right)\,,
\nonumber\\
&&O_2 = \bar{q} \left(-iv\cdot\overleftarrow{D}\right) h_v
= - i v\cdot\partial \left(\bar{q}h_v\right)\,,
\nonumber\\
&&O_3 = i \int dx\,T\left\{\tilde{\jmath}(0),O_k(x)\right\}\,,
\nonumber\\
&&O_4 = i \int dx\,T\left\{\tilde{\jmath}(0),O_m(x)\right\}\,.
\label{O4}
\end{eqnarray}
From~(\ref{Biloc}) and~(\ref{Repar}),
\begin{eqnarray}
&&B^1_1 = B^1_3 = C_1\,,\quad
B^1_4 = C_m C_1\,,
\nonumber\\
&&B^\V_1 = C_\V - 2 C_{\gamma_\bot}\,,\quad
B^\V_3 = C_\V\,,\quad
B^\V_4 = C_m C_\V\,.
\label{O134}
\end{eqnarray}
The anomalous dimension matrix of the dimension-4 operators~(\ref{O4})
has the structure~\cite{Ne:94}
\begin{equation}
\tilde{\gamma} +
\left(
\begin{array}{cccc}
0 & 0        & 0 & 0 \\
0 & 0        & 0 & 0 \\
0 & \gamma^k & 0 & 0 \\
0 & \gamma^m & 0 & \gamma_m
\end{array}
\right)\,.
\label{gamma4}
\end{equation}
The operators $O_{1,2}$ are renormalized multiplicatively with $\tilde{\gamma}$
(this fixes the first two rows);
the form of $B_{1,3,4}$~(\ref{O134}) fixes the columns 1, 3, 4.
The mixing of the local operator $O_2$ into the bilocal operators $O_{3,4}$
is described by~\cite{BNP:01}
(these anomalous dimensions are called $\gamma^{\mathrm{kin}}_1$
and $\gamma^{\mathrm{mag}}_1+\gamma^{\mathrm{mag}}_2+\gamma^{\mathrm{mag}}_3$
in this paper)
\begin{eqnarray}
&&\gamma^k = - 8 C_F \frac{\alpha_s}{4\pi}
+ C_F \Biggl[ \left(-\frac{32}{9}\pi^2-\frac{32}{3}\right) C_F
\nonumber\\
&&\qquad{} + \left(\frac{16}{3}\pi^2-\frac{608}{9}\right) C_A
+ \frac{160}{9} T_F n_l \Biggr]
\left(\frac{\alpha_s}{4\pi}\right)^2 + \cdots
\nonumber\\
&&\gamma^m = 8 C_F \frac{\alpha_s}{4\pi}
+ C_F \Biggl[ \left(\frac{128}{9}\pi^2+\frac{32}{3}\right) C_F
\nonumber\\
&&\qquad{} + \left(-\frac{32}{9}\pi^2+\frac{548}{9}\right) C_A
- \frac{160}{9} T_F n_l \Biggr]
\left(\frac{\alpha_s}{4\pi}\right)^2 + \cdots
\label{Mix}
\end{eqnarray}

The unknown coefficients $B^\Gamma_2(\mu',\mu)$ for $\Gamma=1$, $\rlap/v$
are obtained by solving the renormalization-group equations
\begin{equation}
\frac{\partial B^\Gamma_2}{\partial\log\mu} =
\tilde{\gamma} B^\Gamma_2
+ \gamma^k B^\Gamma_3
+ \gamma^m B^\Gamma_4\,,
\label{RGB2}
\end{equation}
with the initial conditions $B^\Gamma_2(m,m)$ obtained by matching at $\mu'=\mu=m$.
The ratio $B^\Gamma_2(\mu',\mu)/C_\Gamma(\mu',\mu)$
does not depend on $\mu'$:
\begin{eqnarray}
&&\frac{B^\Gamma_2(\mu',\mu)}{C_\Gamma(\mu',\mu)} =
\frac{\hat{B}^\Gamma_2}{\hat{C}_\Gamma}
- \int_{\alpha_s(\mu_0)}^{\alpha_s(\mu)} \frac{\gamma^k(\alpha_s)}{2\beta(\alpha_s)}
\frac{d\alpha_s}{\alpha_s}
\nonumber\\
&&\quad{} - \hat{C}_m \alpha_s(\mu_0)^{\frac{\gamma_{m0}}{2\beta_0}}
\int_{\alpha_s(\mu_0)}^{\alpha_s(\mu)} \frac{\gamma^m(\alpha_s)}{2\beta(\alpha_s)}
K_{-\gamma_m}(\alpha_s) \alpha_s^{-\frac{\gamma_{m0}}{2\beta_0}}
\frac{d\alpha_s}{\alpha_s}
\label{RGB2sol}
\end{eqnarray}
(see~(\ref{RGCm})).
The renormalization-group invariants $\hat{B}^\Gamma_2$ start from one loop:
$\hat{B}^\Gamma_2=b^\Gamma_{21}\alpha_s(\mu_0)/(4\pi)+\cdots$.

To find $B^\Gamma_2(m,m)$ for $\Gamma=1$, $\rlap/v$ with one-loop accuracy,
we write down the sum in~(\ref{Exp}) via the bare operators:
\begin{eqnarray*}
&&C_F \frac{g_0^2 m^{-2\varepsilon}}{(4\pi)^{d/2}} \Gamma(\varepsilon) b_\Gamma O_{20}
\pm O_{10} + O_{30} + O_{40}\\
&&{} = \left( C_F b_\Gamma - \frac{\gamma^k_0+\gamma^m_0}{2} \right)
\frac{\alpha_s(m)}{4\pi\varepsilon} O_2(m)
+ (\mbox{other operators})\,.
\end{eqnarray*}
Taking $\gamma^k_0+\gamma^m_0=0$ into account, we see that both $b_\Gamma$
should vanish at $\varepsilon=0$.
The $\mathcal{O}(\varepsilon)$ terms of~(\ref{Vert1}) give~\cite{GH:91,Ne:94,Ba:94}
\[
B^1_2(m,m) = 8 C_F \frac{\alpha_s(m)}{4\pi} + \cdots\,,\quad
B^\V_2(m,m) = 12 C_F \frac{\alpha_s(m)}{4\pi} + \cdots\,,
\]
and
\begin{equation}
\hat{B}^1_2 = 8 C_F \frac{\alpha_s(\mu_0)}{4\pi} + \cdots\,,\quad
\hat{B}^\V_2 = 12 C_F \frac{\alpha_s(\mu_0)}{4\pi} + \cdots
\label{B2hat1}
\end{equation}
(using $\gamma^k_0+\gamma^m_0=0$).

An exact relation between $\hat{B}^1_2$ and $\hat{B}^\V_2$ can be derived.
The QCD vector current and the scalar one are related by the equations of motion:
\begin{equation}
i \partial_\alpha j_0^\alpha = i \partial_\alpha j^\alpha
= m_0 j_0 = \bar{m}(\mu') j(\mu')\,,
\label{Div}
\end{equation}
where
\[
\bar{m}(\mu') = \hat{m}
\left(\frac{\alpha_s'(\mu')}{\alpha_s'(\mu_0)}\right)^{-\frac{\gamma'_{10}}{2\beta'_0}}
K'_{-\gamma'_1}(\alpha_s'(\mu'))
\]
is the $n_f$-flavour $\overline{\mathrm{MS}}$ running mass,
and~(\ref{gammaG}) $\gamma'_1=-6C_F\alpha_s/(4\pi)+\cdots$ is minus the mass
anomalous dimension.
We separate $j^\alpha=(j\cdot v)v^\alpha+j_\bot^\alpha$
and substitute the expansions~(\ref{Exp}) with~(\ref{O4}), (\ref{O134}).
The matrix element of~(\ref{Div}) from the heavy quark with momentum $mv$
to the on-shell light quark with momentum $p$ reads
\begin{eqnarray*}
&&m C_\V(\mu',\mu) \left\{ 1 + \frac{1}{2m}
\left[ \left(\frac{B^\V_2(\mu',\mu)}{C_\V(\mu',\mu)}
+2\left(\frac{C_{\gamma_\bot}(\mu',\mu)}{C_\V(\mu',\mu)}-1\right)\right)p\cdot v
+ r \right] \right\}\\
&&{} = \bar{m}(\mu') C_1(\mu',\mu) \left\{ 1 + \frac{1}{2m}
\left[ \frac{B^1_2(\mu',\mu)}{C_1(\mu',\mu)} p\cdot v + r \right]
\right\}\,,
\end{eqnarray*}
where
\[
r = \frac{{<}q|i\int dx\,T\left\{\tilde{\jmath}(\mu),
\left(O_k+C_m(\mu)O_m(\mu)\right)_x\right\}|Q{>}
-p_\alpha\,{<}q|\tilde{\jmath}^\alpha(\mu)|Q{>}}
{{<}q|\tilde{\jmath}(\mu)|Q{>}}\,,
\]
$\tilde{\jmath}^\alpha=\bar{q}\gamma^\alpha h_v$.
At the leading order in $1/m$, this yields~\cite{BG:95}
\begin{equation}
\frac{m}{\bar{m}(\mu')} = \frac{C_1(\mu',\mu)}{C_\V(\mu',\mu)}
\quad\mbox{or}\quad
\frac{m}{\hat{m}} = \frac{\hat{C}_1}{\hat{C}_\V}\,.
\label{mm0}
\end{equation}
At the first order, we obtain
\begin{eqnarray}
&&\frac{B^1_2(\mu',\mu)}{C_1(\mu',\mu)} - \frac{B^\V_2(\mu',\mu)}{C_\V(\mu',\mu)} =
2 \left( \frac{C_{\gamma_\bot}(\mu',\mu)}{C_\V(\mu',\mu)} - 1 \right)
\nonumber\\
&&\mbox{or}\quad
\frac{\hat{B}^1_2}{\hat{C}_1} - \frac{\hat{B}^\V_2}{\hat{C}_\V} =
2 \left( \frac{\hat{C}_{\gamma_\bot}}{\hat{C}_\V} - 1 \right)\,.
\label{mm1}
\end{eqnarray}
Note that~(\ref{Ward1}) is just the one-loop case of this general result.
The one-loop results~(\ref{B2hat1}), of course,
satisfy this requirement.

As discussed above, these results don't change if we replace
$\bar{q}\to\bar{q}\gamma_5^{\mathrm{AC}}$.
Now the leading term is $\tilde{\jmath}=\bar{q}\gamma_5^{\mathrm{AC}}h_v$,
and the definitions of $O_i$~(\ref{O4}) are changed accordingly.
We define the matrix elements
\begin{eqnarray}
&&{<}0|\left(\bar{q}\gamma_5^{\mathrm{AC}}Q\right)_{\mu'}|B{>} =
- i m_B f^P_B(\mu')\,,
\nonumber\\
&&{<}0|\bar{q}\gamma_5^{\mathrm{AC}}\gamma^\alpha Q|B{>} =
- i f_B p_B^\alpha\,,
\label{Qmat1}
\end{eqnarray}
where
\begin{eqnarray}
&&f^P_B(\mu') = \hat{f}^P_B
\left(\frac{\alpha_s'(\mu')}{\alpha_s'(\mu_0)}\right)^{\frac{\gamma'_{10}}{2\beta'_0}}
K'_{\gamma'_1}(\alpha_s'(\mu'))\,.
\label{fPB}
\end{eqnarray}

The HQET matrix elements are~\cite{Ne:92}
\begin{eqnarray}
&&{<}0|\tilde{\jmath}(\mu)|B{>} = - i \sqrt{m_B} F(\mu)\,,
\nonumber\\
&&{<}0|O_1(\mu)|B{>} = - {<}0|O_2|B{>} = - i \sqrt{m_B} \bar{\Lambda} F(\mu)\,,
\nonumber\\
&&{<}0|O_3(\mu)|B{>} = - i \sqrt{m_B} F(\mu) G_k(\mu)\,,
\nonumber\\
&&{<}0|O_4(\mu)|B{>} = - i \sqrt{m_B} F(\mu) G_m(\mu)
\label{Hmat1}
\end{eqnarray}
(in~\cite{Ne:92}, $G_k$ and $G_m$ are called $2G_1$ and $12G_2$),
where $\bar{\Lambda}=m_B-m$ is the $B$-meson residual energy,
\begin{equation}
F(\mu) = \hat{F}
\left(\frac{\alpha_s(\mu)}{4\pi}\right)^{\frac{\tilde{\gamma}_0}{2\beta_0}}
K_{\tilde{\gamma}}(\alpha_s(\mu))\,,
\label{Fmu}
\end{equation}
and $\hat{F}$ is $\mu$-independent and is thus 
just a (non-perturbative) number times $\Lambda_{\overline{\mathrm{MS}}}^{3/2}$.
The hadronic parameters $G_{k,m}(\mu)$ obey the renormalization-group equations
\begin{eqnarray}
&&\frac{d G_k(\mu)}{d\log\mu} = \gamma^k(\alpha_s(\mu)) \bar{\Lambda}\,,
\nonumber\\
&&\frac{d G_m(\mu)}{d\log\mu} + \gamma_m(\alpha_s(\mu)) G_m(\mu) =
\gamma^m(\alpha_s(\mu)) \bar{\Lambda}\,.
\label{RGG}
\end{eqnarray}
Their solution is
\begin{eqnarray}
&&G_k(\mu) = \hat{G}_k - \bar{\Lambda}
\Biggl[ \frac{\gamma^k_0}{2\beta_0} \log \frac{\alpha_s(\mu)}{4\pi}
+ \int_0^{\alpha_s(\mu)}
\left(\frac{\gamma^k(\alpha_s)}{2\beta(\alpha_s)}-\frac{\gamma^k_0}{2\beta_0}\right)
\frac{d\alpha_s}{\alpha_s} \Biggr]\,,
\nonumber\\
&&C_m(\mu) G_m(\mu) = \hat{C}_m
\left(\frac{\alpha_s(\mu_0)}{4\pi}\right)^{\frac{\gamma_{m0}}{2\beta_0}}
\nonumber\\
&&\quad{}\times\Biggl[ \hat{G}_m
- \bar{\Lambda} \int_0^{\alpha_s(\mu)}
\frac{\gamma^m(\alpha_s)}{2\beta(\alpha_s)} K_{-\gamma_m}(\alpha_s)
\left(\frac{\alpha_s}{4\pi}\right)^{-\frac{\gamma_{m0}}{2\beta_0}}
\frac{d\alpha_s}{\alpha_s} \Biggr]\,,
\label{RGk}
\end{eqnarray}
where $\hat{G}_k$ and $\hat{G}_m$ are
again $\mu$-independent and thus are just some (non-perturbative) numbers
times $\Lambda_{\overline{\mathrm{MS}}}$.

Taking the matrix element of~(\ref{Exp}), we obtain
\begin{eqnarray}
&&\left\{\begin{array}{c}f^P_B(\mu')\\f_B\end{array}\right\} =
\frac{C_\Gamma(\mu',\mu) F(\mu)}{\sqrt{m_B}}
\nonumber\\
&&\quad{} \times \left[ 1 + \frac{1}{2m}
\left( C^\Gamma_\Lambda(\mu) \bar{\Lambda} + G_k(\mu) + C_m(\mu) G_m(\mu) \right)
\right]\,,
\label{fB1}
\end{eqnarray}
where $\Gamma=1$, $\rlap/v$, and
\[
C^1_\Lambda(\mu) = 1 - \frac{B^1_2(\mu',\mu)}{C_1(\mu',\mu)}\,,\quad
C^\V_\Lambda(\mu) = 1 - 2 \frac{C_{\gamma_\bot}(\mu',\mu)}{C_\V(\mu',\mu)}
- \frac{B^\V_2(\mu',\mu)}{C_\V(\mu',\mu)}\,.
\]
Substituting the solutions of the renormalization-group equations,
we arrive at the explicitly $\mu$-independent expressions
\begin{eqnarray}
&&\left\{\begin{array}{c}\hat{f}^P_B\\f_B\end{array}\right\} =
\left(\frac{\alpha_s(\mu_0)}{4\pi}\right)^{\frac{\tilde{\gamma}_0}{2\beta_0}}
\frac{\hat{C}_\Gamma \hat{F}}{\sqrt{m_B}}
\nonumber\\
&&\quad{}\times\left[ 1 + \frac{1}{2m} \left(
\hat{C}^\Gamma_\Lambda \bar{\Lambda} + \hat{G}_k + \hat{C}_m \hat{G}_m
\left(\frac{\alpha_s(\mu_0)}{4\pi}\right)^{\frac{\gamma_{m0}}{2\beta_0}}
\right) \right]\,,
\label{fB2}
\end{eqnarray}
where
\begin{eqnarray*}
&&\hat{C}^\Gamma_\Lambda = 1
- 2 \left\{\begin{array}{c}0\\\hat{C}_{\gamma_\bot}/\hat{C}_\V\end{array}\right\}
- \frac{\hat{B}^\Gamma_2}{\hat{C}_\Gamma}\\
&&\quad{} - \frac{\gamma^k_0}{2\beta_0} \log \frac{\alpha_s(\mu_0)}{4\pi}
- \int_0^{\alpha_s(\mu_0)}
\left(\frac{\gamma^k(\alpha_s)}{2\beta(\alpha_s)}-\frac{\gamma^k_0}{2\beta_0}\right)
\frac{d\alpha_s}{\alpha_s}\\
&&\quad{} - \hat{C}_m \alpha_s(\mu_0)^{\frac{\gamma_{m0}}{2\beta_0}}
\int_0^{\alpha_s(\mu_0)} \frac{\gamma^m(\alpha_s)}{2\beta(\alpha_s)}
K_{-\gamma_m}(\alpha_s)
\alpha_s^{-\frac{\gamma_{m0}}{2\beta_0}} \frac{d\alpha_s}{\alpha_s}\,.
\end{eqnarray*}
At the next-to-leading order,
\begin{eqnarray}
&&\left\{\begin{array}{c}\hat{f}^P_B\\f_B\end{array}\right\} =
\left(\frac{\alpha_s(\mu_0)}{4\pi}\right)^{\frac{\tilde{\gamma}_0}{2\beta_0}}
\frac{\hat{C}_\Gamma\hat{F}}{\sqrt{m_B}} \Biggl\{ 1
\nonumber\\
&&\quad{} + \frac{1}{2m} \Biggl[
\left(-\frac{\gamma^k_0}{2\beta_0} \log \frac{\alpha_s(\mu_0)}{4\pi}
\pm 1 + \frac{\gamma^m_0}{\gamma_{m0}}
+ c^\Gamma_{\Lambda1} \frac{\alpha_s(\mu_0)}{4\pi} + \cdots \right)
\bar{\Lambda}
\nonumber\\
&&\qquad{} + \hat{G}_k + \hat{G}_m
\left(\frac{\alpha_s(\mu_0)}{4\pi}\right)^{\frac{\gamma_{m0}}{2\beta_0}}
\left(1 + c_{m1} \frac{\alpha_s(\mu_0)}{4\pi} + \cdots\right)
\Biggr] \Biggr\}\,,
\label{fB3}
\end{eqnarray}
where
\begin{eqnarray*}
&&c^\Gamma_{\Lambda1} = (1\mp1) \left(c^\V_1-c^{\gamma_\bot}_1\right)
- b^\Gamma_{21}
+ \frac{\gamma^m_0}{\gamma_{m0}} c_{m1}
- \frac{\gamma^k_1+\gamma^m_1}{2\beta_0}
+ \frac{\beta_1\left(\gamma^k_0+\gamma^m_0\right)}{2\beta_0^2}\\
&&\quad{}
+ \frac{\gamma_{m0}\gamma^m_1-\gamma^m_0\gamma_{m1}}
{2\beta_0\left(\gamma_{m0}-2\beta_0\right)}
\end{eqnarray*}
(here $\gamma^k_0+\gamma^m_0=0$).

%As already discussed in Sect.~\ref{Intro},
%the meson matrix elements are (in principle) measurable quantities,
%and should be unambiguous.
%In the $1/m$ expansion~(\ref{fB1}),
%they are expressed via short-distance Wilson coefficients
%and long-distance hadronic matrix elements.
%In regularization schemes without a hard momentum cut-off,
%Wilson coefficients also contain contributions from large distances,
%where perturbation theory is ill defined,
%and these contributions produce infrared renormalon ambiguities.
%Likewise, hadronic matrix elements contain contributions from small distances,
%which lead to ultraviolet renormalon ambiguities.
%Only when all contributions are combined to form a physical quantity,
%an unambiguous result is obtained~\cite{NS:95}.

In the large $\beta_0$ limit,
the leading short-distance coefficient $\hat{C}_\Gamma$ in~(\ref{fB3})
has the IR renormalon ambiguity~(\ref{DCLb0}).
If we change the prescription for calculating the integral~(\ref{Laplace})
near $u=\frac{1}{2}$, we should adjust the hadronic parameters $\bar{\Lambda}$,
$\hat{G}_k$, $\hat{G}_m$ accordingly.
The UV renormalon ambiguity of $\bar{\Lambda}$ is given by~(\ref{DLbar})
in this limit.
The UV renormalon ambiguities of $G_{k,m}(\mu)$~\cite{NS:95}
\begin{equation}
\Delta G_k(\mu) = - \frac{3}{2} \Delta \bar{\Lambda}\,,\quad
\Delta G_m(\mu) = 2 \Delta \bar{\Lambda}
\label{DGkm}
\end{equation}
are $\mu$-independent.
We can obtain them by a direct calculation, performing
a similar analysis as in~\cite{GN:97}, see Appendix~\ref{UV}.
In the large $\beta_0$ limit, we obtain from~(\ref{RGk}) and~(\ref{DGkm})
\begin{equation}
\Delta \hat{G}_k = - \frac{3}{2} \Delta \bar{\Lambda}\,,\quad
\Delta \hat{G}_m = \left(2-\frac{\gamma^m_0}{\gamma_{m0}}\right)
\Delta \bar{\Lambda}\,.
\label{DGhat}
\end{equation}
In the $1/m$ expansion for the meson decay constants~(\ref{fB3}),
$\Gamma=1$ and $\rlap/v$ correspond to
$n=0$, $\eta=-1$ and $n=1$, $\eta=1$ in~(\ref{DCLb0}),
and the IR renormalon ambiguities~(\ref{DCLb0}) are
$-(3/4)(\Delta\bar{\Lambda}/m)$ and $(1/4)(\Delta\bar{\Lambda}/m)$.
They cancel with the UV renormalon ambiguities
of the hadronic matrix elements~(\ref{DLbar}), (\ref{DGhat})
in the $1/m$ correction in~(\ref{fB3}).
In fact, the results~(\ref{DGkm}) were first obtained~\cite{NS:95}
from the \emph{requirement} of such cancellation for $\Gamma=\rlap/v$,
$\gamma_\bot$, by solving a system of two linear equations,
and later confirmed~\cite{BG:95} by considering all possible $\Gamma$
(this gives three equations, thus providing a consistency check).
Here we demonstrate the cancellation of renormalon ambiguities
by a direct calculation.

This cancellation should hold also beyond the large $\beta_0$ limit.
By dimensional arguments, the UV renormalon ambiguities of
the $\mu$-independent
hadronic parameters $\bar{\Lambda}$, $\hat{G}_k$, $\hat{G}_m$
must be equal to $\Lambda_{\overline{\mathrm{MS}}}$ times some numbers:
\begin{eqnarray}
&&\Delta \bar{\Lambda} = N_0 \Delta_0\,,\quad
\Delta \hat{G}_k = - \frac{3}{2} N_1 \Delta_0\,,\quad
\Delta \hat{G}_m = N_2 \left(2-\frac{\gamma^m_0}{\gamma_{m0}}\right) \Delta_0\,,
\nonumber\\
&&\Delta_0 = - 2 C_F e^{5/6} \frac{\Lambda_{\overline{\mathrm{MS}}}}{\beta_0}\,.
\label{Dbeyond}
\end{eqnarray}
The normalization factors are unity in the large $\beta_0$ limit,
$N_i=1+\mathcal{O}(1/\beta_0)$; in general, they are just some
unknown numbers of order one.
Using
\begin{eqnarray}
&&\Lambda_{\overline{\mathrm{MS}}} =
\mu_0 \exp \left[-\frac{2\pi}{\beta_0\alpha_s(\mu_0)}\right]
\left(\frac{\alpha_s(\mu_0)}{4\pi}\right)^{-\frac{\beta_1}{2\beta_0^2}}
K(\alpha_s(\mu_0))\,,
\label{LambdaV}\\
&&K(\alpha_s) =
\exp \int_0^{\alpha_s} \left( \frac{1}{2\beta(\alpha_s)}
- \frac{2\pi}{\beta_0\alpha_s} + \frac{\beta_1}{2\beta_0^2} \right)
\frac{d\alpha_s}{\alpha_s} =
1 - \frac{\beta_0\beta_2-\beta_1^2}{2\beta_0^3} \frac{\alpha_s}{4\pi}
+ \cdots\,,
\nonumber
\end{eqnarray}
$\mu_0=e^{-5/6}m$~(\ref{mu0}),
we can represent the UV renormalon ambiguity of the $1/m$ correction
in~(\ref{fB3}) as $\exp[-2\pi/(\beta_0\alpha_s(\mu_0))]$
times a sum of terms with different fractional powers of $\alpha_s(\mu_0)/(4\pi)$.%
\footnote{It is convenient to replace
$\log[\alpha_s(\mu_0)/(4\pi)]\to [(\alpha_s(\mu_0)/(4\pi))^\delta-1]/\delta$,
and take the limit $\delta\to 0$ at the end of calculation.}
In order to cancel this ambiguity, we should have the branching point
\begin{equation}
S_\Gamma(u) = \sum_i \frac{r_i}{\left(\frac{1}{2}-u\right)^{1+a_i}}
+ S_\Gamma^{\mathrm{reg}}(u)
\label{Branching}
\end{equation}
instead of a simple pole~(\ref{SuLb0})
(here $S_\Gamma^{\mathrm{reg}}(u)$ is regular at $u=\frac{1}{2}$).
We define the IR renormalon ambiguity of $\hat{C}_\Gamma$ -- generalizing
the prescription to take the residue of the pole -- to be
the integral of~(\ref{Branching}) around the cut divided by $2\pi i$:
\begin{equation}
\Delta \hat{C}_\Gamma = \frac{1}{\beta_0}
\exp\left[-\frac{2\pi}{\beta_0\alpha_s(\mu_0)}\right]
\sum_i \frac{r_i}{\Gamma(1+a_i)}
\left(\frac{\beta_0\alpha_s(\mu_0)}{4\pi}\right)^{-a_i}\,.
\label{DCgen}
\end{equation}
The requirement of cancellation of the ambiguities gives
\begin{eqnarray}
&&S_\Gamma(u) =
\frac{C_F}{\left(\frac{1}{2}-u\right)^{1+\frac{\beta_1}{2\beta_0^2}}}
\left[1+c^{\Gamma\prime}_1\left(\tfrac{1}{2}-u\right)+\cdots\right]
\Biggl\{
\nonumber\\
&&\left[ - \frac{\gamma^k_0}{2\beta_0}
\left( \log \frac{\frac{1}{2}-u}{\beta_0}
- \psi\left(1+\frac{\beta_1}{2\beta_0^2}\right) \right)
\pm 1 + \frac{\gamma^m_0}{\gamma_{m0}}
+ c^{\Gamma\prime}_{\Lambda1} \left(\tfrac{1}{2}-u\right) + \cdots
\right] N_0'
\nonumber\\
&&{} - \frac{3}{2} N_1'
+ \left(2-\frac{\gamma^m_0}{\gamma_{m0}}\right) N_2'
\left(\tfrac{1}{2}-u\right)^{\frac{\gamma_{m0}}{2\beta_0}}
\left[1 + c_{m1}'\left(\tfrac{1}{2}-u\right) + \cdots\right]
\Biggr\}\,,
\label{Su1}
\end{eqnarray}
where
\begin{eqnarray}
&&\frac{N_0'}{N_0} = \frac{N_1'}{N_1} =
\Gamma\left(1+\frac{\beta_1}{2\beta_0^2}\right)
\beta_0^{\frac{\beta_1}{2\beta_0^2}}\,,\quad
\frac{N_2'}{N_2} =
\Gamma\left(1+\frac{\beta_1}{2\beta_0^2}-\frac{\gamma_{m0}}{2\beta_0}\right)
\beta_0^{\frac{\beta_1}{2\beta_0^2}-\frac{\gamma_{m0}}{2\beta_0}}\,,
\nonumber\\
&&c^{\Gamma\prime}_1 = \frac{2\beta_0}{\beta_1}
\left(c^\Gamma_1 - \frac{\beta_0\beta_2-\beta_1^2}{2\beta_0^3}\right)\,,\quad
c^{\Gamma\prime}_{\Lambda1} = \frac{2\beta_0}{\beta_1}
\left( c^\Gamma_{\Lambda1} - \tfrac{1}{2} \gamma^k_0 c^{\Gamma\prime}_1 \right)\,,
\nonumber\\
&&c_{m1}' = \frac{\beta_0\left(2c_{m1}+\gamma_{m0}c^{\Gamma\prime}_1\right)}
{\beta_1-\beta_0\gamma_{m0}}\,.
\label{Npr}
\end{eqnarray}
In the large $\beta_0$ limit, the formula~(\ref{Su1})
reproduces the known results~(\ref{SuLb0}), (\ref{DCLb0}).

The asymptotics of $c^\Gamma_L$~(\ref{series}) at $L\gg1$
is determined by the renormalon singularity closest to the origin
(see~(\ref{Borel})).
At $n\gg1$,
\[
\Gamma(n+a+1) = n!\,n^a\,\left(1 + \frac{a(a+1)}{2n}
+ \frac{a(a^2-1)(3a+2)}{24n^2} + \cdots\right)\,,
\]
and we arrive at
\begin{eqnarray}
&&c^\Gamma_{n+1} = 2 C_F\,n!\,(2\beta_0)^n\,(2 \beta_0 n)^{\frac{\beta_1}{2\beta_0^2}}
\left(1 + \frac{c^{\Gamma\prime\prime}_1}{2 \beta_0 n} + \cdots\right)
\Biggl[
\nonumber\\
&&\quad\left( \frac{\gamma^k_0}{2\beta_0} \log 2 \beta_0 n
\pm 1 + \frac{\gamma^m_0}{\gamma_{m0}}
+ \frac{c^{\Gamma\prime\prime}_{\Lambda1}}{2 \beta_0 n}
+ \cdots \right) N_0
\nonumber\\
&&\quad{} - \frac{3}{2} N_1
+ \left(2-\frac{\gamma^m_0}{\gamma_{m0}}\right) N_2
(2 \beta_0 n)^{-\frac{\gamma_{m0}}{2\beta_0}}
\left(1 + \frac{c_{m1}''}{2 \beta_0 n} + \cdots
\right) \Biggr]\,,
\label{asy1}
\end{eqnarray}
where
\begin{eqnarray}
&&c^{\Gamma\prime\prime}_1 = c^\Gamma_1
- \frac{2\beta_0\beta_2-2\beta_0^2\beta_1-3\beta_1^2}{4\beta_0^3}\,,\quad
c^{\Gamma\prime\prime}_{\Lambda1} = c^\Gamma_{\Lambda1}
+ \gamma^k_0 \frac{\beta_1+\beta_0^2}{2\beta_0^2}\,,
\nonumber\\
&&c_{m1}'' = c_{m1}
- \gamma_{m0} \frac{2(\beta_1+\beta_0^2)-\beta_0\gamma_{m0}}{4\beta_0^2}\,.
\label{Nprpr}
\end{eqnarray}
The result~(\ref{asy1}) is model-independent and the powers of $n$ are exact.
However, the normalization factors $N_i$ cannot be determined within this approach.
The leading term at $n\to\infty$ formally is the logarithmic term,
because $\gamma_{m0}$ is positive (see~(\ref{gammam})).
At moderate values of $n$, all leading terms are of similar importance.
The function $S_\Gamma^{\mathrm{reg}}(u)$ in~(\ref{Branching})
has singularities at $u=1$ (IR) and $u=-1$ (UV),
and thus gives exponentially smaller contributions
with $(\pm\beta_0)^n$ instead of $(2\beta_0)^n$.

The matrix element of~(\ref{Div}) is
\begin{equation}
\frac{f^P_B(\mu')}{f_B} = \frac{m_B}{\bar{m}(\mu')}
\quad\mbox{or}\quad
\frac{\hat{f}^P_B}{f_B} = \frac{m_B}{\hat{m}}\,.
\label{mm2}
\end{equation}
Substituting~(\ref{fB1}) (or~(\ref{fB2})) and using the relation~(\ref{mm1}),
we obtain
\begin{equation}
\frac{f^P_B(\mu')}{f_B} = \frac{C_1(\mu',\mu)}{C_\V(\mu',\mu)}
\left(1 + \frac{\bar{\Lambda}}{m}\right)
\quad\mbox{or}\quad
\frac{\hat{f}^P_B}{f_B} = \frac{\hat{C}_1}{\hat{C}_\V}
\left(1 + \frac{\bar{\Lambda}}{m}\right)\,.
\label{mm3}
\end{equation}
The ratio of the quark masses is given by~(\ref{mm0}).
Naturally, it contains no $1/m$ corrections with $B$-meson hadronic parameters;
it is just a series in $\alpha_s(\mu_0)$, see the Appendix~\ref{MM}.
The Borel image of this series is
\begin{eqnarray}
&&S_{m/\hat{m}}(u) =
\frac{2 C_F N_0'}{\left(\frac{1}{2}-u\right)^{1+\frac{\beta_1}{2\beta_0^2}}}
\left[1 + c^{(m/\hat{m})\prime}_1\left(\tfrac{1}{2}-u\right)
+ c^{(m/\hat{m})\prime}_2\left(\tfrac{1}{2}-u\right)^2
+ \cdots\right]\,,
\nonumber\\
&&c^{(m/\hat{m})\prime}_1 = \frac{2\beta_0}{\beta_1}
\left(c^{(m/\hat{m})}_1 - \frac{\beta_0\beta_2-\beta_1^2}{2\beta_0^3}\right)\,,
\label{Summ}\\
&&c^{(m/\hat{m})\prime}_2 = \frac{4\beta_0^2}{\beta_1(\beta_1-2\beta_0^2)}
\Biggl[ c^{(m/\hat{m})}_2
- \frac{\beta_0\beta_2-\beta_1^2}{2\beta_0^3} c^{(m/\hat{m})}_1
\nonumber\\
&&\quad{} - \frac{2\beta_0^4\beta_3 - \beta_0^2\beta_2^2
+ \beta_1 (\beta_1-2\beta_0^2) (2\beta_0\beta_2-\beta_1^2)}{8\beta_0^6}
\Biggr]\,,
\nonumber
\end{eqnarray}
and the asymptotics of $c^{(m/\hat{m})}_L$ at $L\gg1$ is
\begin{eqnarray}
&&c^{(m/\hat{m})}_{n+1} =
4 C_F N_0\,n!\,(2\beta_0)^n\,(2 \beta_0 n)^{\frac{\beta_1}{2\beta_0^2}}
\left(1 + \frac{c^{(m/\hat{m})\prime\prime}_1}{2\beta_0 n}
+ \frac{c^{(m/\hat{m})\prime\prime}_2}{(2\beta_0 n)^2}
+ \cdots\right)\,,
\nonumber\\
&&c^{(m/\hat{m})\prime\prime}_1 = c^{(m/\hat{m})}_1
- \frac{2\beta_0\beta_2-2\beta_0^2\beta_1-3\beta_1^2}{4\beta_0^3}\,,
\label{cnmm}\\
&&c^{(m/\hat{m})\prime\prime}_2 = c^{(m/\hat{m})}_2
- \frac{2\beta_0\beta_2+2\beta_0^2\beta_1-3\beta_1^2}{4\beta_0^3}
c^{(m/\hat{m})}_1
\nonumber\\
&&\quad{} - \frac{24\beta_0^4\beta_3 - 12\beta_0^2\beta_2^2
+ \beta_1 (\beta_1-2\beta_0^2)
(36\beta_0\beta_2-27\beta_1^2-10\beta_0^2\beta_1-8\beta_0^4)}%
{96\beta_0^6}\,.
\nonumber
\end{eqnarray}
These results are equivalent to~\cite{Be:95}.

We can try to estimate the unknown normalization constant $N_0$
following~\cite{Le:97,Pi:01}.
The function
\[
\tilde{S}_{m/\hat{m}}(u) = (1-2u)^{1+\frac{\beta_1}{2\beta_0^2}} S_{m/\hat{m}}(u)\,,
\quad
S_{m/\hat{m}}(u) = \sum_{L=1}^\infty \frac{c^{(m/\hat{m})}_L}{(L-1)!}
\left(\frac{u}{\beta_0}\right)^{L-1}
\]
still has a singularity at $u=\frac{1}{2}$
due to $S_\Gamma^{\mathrm{reg}}(u)$ in~(\ref{Branching}),
but has a finite limit at $u\to\frac{1}{2}-0$.
The radius of convergence of its expansion in $u$ is thus $\frac{1}{2}$,
but the series should converge at $u=\frac{1}{2}$.
Therefore, we can calculate $\tilde{S}_{m/\hat{m}}\left(\frac{1}{2}\right)$,
and hence $N_0$~(\ref{Summ}),
from this expansion.
Substituting $c^{(m/\hat{m})}_L$ for $L\le3$ from Appendix~\ref{MM}, we find
\begin{equation}
N_0 = 0.288 \cdot (1 + 0.075 + 0.630 + \cdots) \approx 0.491\,.
\label{pineda}
\end{equation}
The three-loop correction turns out to be large,
which casts some doubt on this estimate of $N_0$.

\section{Spin-1 currents}
\label{S1}

In the case of the currents with
$\Gamma=\gamma_\bot^\alpha$, $\gamma_\bot^\alpha\rlap/v$,
the leading term in the expansion~(\ref{Exp}) contains
$\tilde{\jmath}^\alpha=\bar{q}\gamma_\bot^\alpha h_v$,
and the $1/m$ correction -- six operators
\begin{eqnarray}
&&O_1^\alpha = \bar{q} iD^\alpha h_v\,,
\nonumber\\
&&O_2^\alpha = \bar{q} i\D \gamma_\bot^\alpha h_v
= i\partial_\beta \left(\bar{q}\gamma^\beta\gamma_\bot^\alpha h_v\right)\,,
\nonumber\\
&&O_3^\alpha = \bar{q} \left(-i\overleftarrow{D}_\bot^\alpha\right) h_v\,,
\nonumber\\
&&O_4^\alpha = \bar{q} \left(-iv\cdot\overleftarrow{D}\right) \gamma_\bot^\alpha h_v
= - i v\cdot\partial \left(\bar{q} \gamma_\bot^\alpha h_v\right)\,,
\nonumber\\
&&O_5^\alpha = i \int dx\,T\left\{\tilde{\jmath}^\alpha(0),O_k(x)\right\}\,,
\nonumber\\
&&O_6^\alpha = i \int dx\,T\left\{\tilde{\jmath}^\alpha(0),O_m(x)\right\}\,.
\label{O6}
\end{eqnarray}
Note that
\begin{equation}
O_1^\alpha - O_3^\alpha = i \partial_\bot^\alpha (\bar{q}h_v)\,.
\label{FullDeriv}
\end{equation}
From~(\ref{Biloc}) and~(\ref{Repar}),
\begin{eqnarray}
&&B^{\gamma_\bot}_1 = 2 C_\V\,,\quad
-B^{\gamma_\bot}_2 = B^{\gamma_\bot}_5 = C_{\gamma_\bot}\,,
\nonumber\\
&&B^{\gamma_\bot}_6 = C_m C_{\gamma_\bot}\,,
\nonumber\\
&& - \tfrac{1}{2}B^{\gamma_\bot\V}_1 = B^{\gamma_\bot\V}_2
= B^{\gamma_\bot\V}_5 = C_{\gamma_\bot\V}\,,
\nonumber\\
&&B^{\gamma_\bot\V}_6 = C_m C_{\gamma_\bot\V}\,.
\label{O1256}
\end{eqnarray}
The anomalous dimension matrix of the dimension-4 operators~(\ref{O6})
has the structure~\cite{Ne:94}
\begin{equation}
\tilde{\gamma} +
\left(
\begin{array}{cccccc}
0 & 0 & \gamma_a   & \gamma_b   & 0 & 0 \\
0 & 0 & 0          & 0          & 0 & 0 \\
0 & 0 & \gamma_a   & \gamma_b   & 0 & 0 \\
0 & 0 & 0          & 0          & 0 & 0 \\
0 & 0 & 0          & \gamma^k   & 0 & 0 \\
0 & 0 & \gamma^m_1 & \gamma^m_2 & 0 & \gamma_m
\end{array}
\right)\,.
\label{gamma6}
\end{equation}
The operators $O_{2,4}^\alpha$ are renormalized
multiplicatively with $\tilde{\gamma}$, which determines 
the second and the fourth row.
The same holds for $O_1^\alpha-O_3^\alpha$~(\ref{FullDeriv}),
and hence the first and the third rows coincide. Furthermore, 
the form of $B_{1,2,5,6}$ (see~(\ref{O1256}))
fixes the columns 1, 2, 5, 6.
Renormalization of the operators~(\ref{O6}) is discussed
in Appendix~\ref{R6} in detail.

The unknown coefficients
$B^\Gamma_{3,4}(\mu',\mu)$ for $\Gamma=\gamma_\bot$, $\gamma_\bot\rlap/v$
are obtained by solving the renormalization-group equations
\begin{eqnarray}
&&\frac{\partial B^\Gamma_3}{\partial\log\mu} =
\left(\tilde{\gamma}+\gamma_a\right) B^\Gamma_3
+ \gamma_a B^\Gamma_1
+ \gamma^m_1 B^\Gamma_6\,,
\nonumber\\
&&\frac{\partial B^\Gamma_4}{\partial\log\mu} =
\tilde{\gamma} B^\Gamma_4
+ \gamma_b \left(B^\Gamma_1+B^\Gamma_3\right)
+ \gamma^k B^\Gamma_5
+ \gamma^m_2 B^\Gamma_6
\label{RGB34}
\end{eqnarray}
(where~(\ref{gammaG}) $\gamma'_{\gamma_\bot\V}=2C_F\alpha_s/(4\pi)+\cdots$)
with the initial conditions $B^\Gamma_{3,4}(m,m)$ obtained by matching.
The ratios $B^\Gamma_i(\mu',\mu)/C_\Gamma(\mu',\mu)$ don't depend on $\mu'$:
\begin{eqnarray}
&&\frac{B^\Gamma_3(\mu',\mu)}{C_\Gamma(\mu',\mu)} =
\left(\frac{\alpha_s(\mu)}{\alpha_s(\mu_0)}\right)^{-\frac{\gamma_{a0}}{2\beta_0}}
K_{-\gamma_a}(\alpha_s(\mu)) \Biggl[ \frac{\hat{B}^\Gamma_3}{\hat{C}_\Gamma}
\nonumber\\\
&&\quad{}
- \left\{\begin{array}{cc}\hat{C}_\V/\hat{C}_{\gamma_\bot}\\-1\end{array}\right\}
\alpha_s(\mu_0)^{-\frac{\gamma_{a0}}{2\beta_0}}
\int_{\alpha_s(\mu_0)}^{\alpha_s(\mu)} \frac{\gamma_a(\alpha_s)}{\beta(\alpha_s)}
K_{\gamma_a}(\alpha_s)
\alpha_s^{\frac{\gamma_{a0}}{2\beta_0}}
\frac{d\alpha_s}{\alpha_s}
\label{B3}\\
&&\quad{} - \hat{C}_m \alpha_s(\mu_0)^{\frac{\gamma_{m0}-\gamma_{a0}}{2\beta_0}}
\int_{\alpha_s(\mu_0)}^{\alpha_s(\mu)} \frac{\gamma^m_1(\alpha_s)}{2\beta(\alpha_s)}
K_{\gamma_a-\gamma_m}(\alpha_s)
\alpha_s^{\frac{\gamma_{a0}-\gamma_{m0}}{2\beta_0}}
\frac{d\alpha_s}{\alpha_s} \Biggr]\,,
\nonumber\\
&&\frac{B^\Gamma_4(\mu',\mu)}{C_\Gamma(\mu',\mu)} =
\frac{\hat{B}^\Gamma_4}{\hat{C}_\Gamma}
- \int_{\alpha_s(\mu_0)}^{\alpha_s(\mu)}
\frac{\gamma_b(\alpha_s)}{2\beta(\alpha_s)}
\left[ \left.\frac{B^\Gamma_3}{C_\Gamma}\right|_{\alpha_s}
+ 2 \left\{\begin{array}{cc}\hat{C}_\V/\hat{C}_{\gamma_\bot}\\-1\end{array}\right\}
\right] \frac{d\alpha_s}{\alpha_s}
\nonumber\\
&&\quad{} - \int_{\alpha_s(\mu_0)}^{\alpha_s(\mu)}
\frac{\gamma^k(\alpha_s)}{2\beta(\alpha_s)} \frac{d\alpha_s}{\alpha_s}
\nonumber\\
&&\quad{} - \hat{C}_m \alpha_s(\mu_0)^{\frac{\gamma_{m0}}{2\beta_0}}
\int_{\alpha_s(\mu_0)}^{\alpha_s(\mu)}
\frac{\gamma^m_2(\alpha_s)}{2\beta(\alpha_s)}
K_{-\gamma_m}(\alpha_s) \alpha_s^{-\frac{\gamma_{m0}}{2\beta_0}}
\frac{d\alpha_s}{\alpha_s}
\label{B4}
\end{eqnarray}
(in the last formula, the running $B^\Gamma_3(\mu',\mu)/C_\Gamma(\mu',\mu)$
corresponding to the integration variable $\alpha_s$ is understood).

To find $B^\Gamma_{3,4}(m,m)$ for $\Gamma=\gamma_\bot$, $\gamma_\bot\rlap/v$
with the one-loop accuracy,
we write down the sum in~(\ref{Exp}) via the bare operators:
\begin{eqnarray*}
&&C_F \frac{g_0^2 m^{-2\varepsilon}}{(4\pi)^{d/2}} \Gamma(\varepsilon)
\left( b_{\Gamma,2} O_{30}^\alpha + b_{\Gamma,1} O_{40}^\alpha \right)
\pm O_{10} \mp O_{20} + O_{50} + O_{60}\\
&&{} = \frac{\alpha_s(m)}{4\pi\varepsilon} \Biggl[
\left( C_F b_{\Gamma,2} \mp \gamma_{a0} - \frac{\gamma^m_{10}}{2} \right)
O_3^\alpha(m)\\
&&\quad{} + \left( C_F b_{\Gamma,1} \mp \gamma_{b0}
- \frac{\gamma^k_0+\gamma^m_{20}}{2} \right) O_4^\alpha(m)
+ (\mbox{other operators}) \Biggr]\,.
\end{eqnarray*}
The values of $b_{\Gamma,i}$ at $\varepsilon=0$ have to cancel
these anomalous dimensions.
The $\mathcal{O}(\varepsilon)$ terms of~(\ref{Vert1}) give
\begin{eqnarray*}
&&B^{\gamma_\bot}_3(m,m) = 4 C_F \frac{\alpha_s(m)}{4\pi} + \cdots\,,\quad
B^{\gamma_\bot}_4(m,m) = - 4 C_F \frac{\alpha_s(m)}{4\pi} + \cdots\,,\\
&&B^{\gamma_\bot\V}_3(m,m) = 2 C_F \frac{\alpha_s(m)}{4\pi} + \cdots\,,\quad
B^{\gamma_\bot\V}_4(m,m) = - 6 C_F \frac{\alpha_s(m)}{4\pi} + \cdots
\end{eqnarray*}
The results for $\Gamma=\gamma_\bot$ were obtained in~\cite{GH:91,Ne:94};
those for $\Gamma=\gamma_\bot\rlap/v$ are new.
Using the one-loop anomalous dimensions, we get
\begin{eqnarray}
&&\hat{B}^{\gamma_\bot}_3 = \frac{2}{3} C_F \frac{\alpha_s(\mu_0)}{4\pi} + \cdots\,,\quad
\hat{B}^{\gamma_\bot}_4 = 6 C_F \frac{\alpha_s(\mu_0)}{4\pi} + \cdots\,,
\nonumber\\
&&\hat{B}^{\gamma_\bot\V}_3 = \frac{26}{3} C_F \frac{\alpha_s(\mu_0)}{4\pi} + \cdots\,,\quad
\hat{B}^{\gamma_\bot\V}_4 = \frac{2}{3} C_F \frac{\alpha_s(\mu_0)}{4\pi} + \cdots
\label{B34hat1}
\end{eqnarray}

We define the matrix elements
\begin{eqnarray}
&&{<}0|\bar{q}\gamma^\alpha Q|B^*{>} = m_{B^*} f_{B^*} e^\alpha\,,
\nonumber\\
&&{<}0|\left(\bar{q}\sigma^{\alpha\beta}Q\right)_{\mu'}|B^*{>} =
i f^T_{B^*}(\mu') \left(e^\alpha p_{B^*}^\beta - e^\beta p_{B^*}^\alpha\right)\,,
\label{Qmat2}\\
&&f^T_{B^*}(\mu') = \hat{f}^T_{B^*}
\left(\frac{\alpha_s'(\mu')}{\alpha_s'(\mu_0)}\right)^{\frac{\gamma'_{\sigma0}}{2\beta'_0}}
K'_{\gamma'_{\sigma}}(\alpha_s'(\mu'))\,,
\nonumber
\end{eqnarray}
where $e^\alpha$ is the $B^*$-meson polarization vector.

The HQET matrix elements are~\cite{Ne:92}
\begin{eqnarray}
&&{<}0|\tilde{\jmath}^\alpha(\mu)|B^*{>} = \sqrt{m_{B^*}} F(\mu) e^\alpha\,,
\label{Hmat2}\\
&&{<}0|O_2^\alpha(\mu)|B^*{>} = {<}0|O_4^\alpha(\mu)|B^*{>} =
- \sqrt{m_{B^*}} \bar{\Lambda} F(\mu) e^\alpha\,.
\nonumber
\end{eqnarray}
The matrix elements of $O_1^\alpha$ and $O_3^\alpha$ are equal,
due to~(\ref{FullDeriv}).
However, the formulae~\cite{Ne:92} for these matrix elements
hold only at the leading order.
Let's define
\begin{equation}
{<}0|O_1^\alpha(\mu)|B^*{>} = {<}0|O_3^\alpha(\mu)|B^*{>} =
- \frac{1}{3} \sqrt{m_{B^*}} \bar{\Lambda} F(\mu) R(\alpha_s(\mu)) e^\alpha\,,
\label{Rdef}
\end{equation}
where $R=1+\mathcal{O}(\alpha_s)$.
It obeys the renormalization-group equation
\begin{equation}
\frac{d\,R}{d\log\mu} + \gamma_a R + 3 \gamma_b = 0\,.
\label{RGR}
\end{equation}
Following~\cite{Ne:92}, we define
\begin{equation}
{<}0|O^\Gamma_1(\mu)|M{>} = \frac{F_2(\mu)}{2} \Tr \gamma_\bot^\alpha \Gamma \mathcal{M}\,,
\label{F2def}
\end{equation}
where $O^\Gamma_1$ is defined in~(\ref{OGamma}),
$M=B$ or $B^*$, and $\mathcal{M}$ is the corresponding Dirac structure.
If we take $\Gamma=\gamma_\alpha \Gamma'$, then
\[
{<}0|O^\Gamma_1(\mu)|M{>} = \frac{3}{2} F_2(\mu) \Tr \Gamma' \mathcal{M}\,.
\]
Taking into account~(\ref{OOr}) and
\[
{<}0|O'_1(\mu)|M{>} = {<}0|O'_2(\mu)|M{>} =
- \frac{1}{2} \bar{\Lambda} F(\mu) \Tr \Gamma' \mathcal{M}\,,
\]
we obtain at the next-to-leading order
\begin{equation}
F_2(\mu) = - \frac{1}{3} \bar{\Lambda} F(\mu) R(\alpha_s(\mu))\,,\quad
R(\alpha_s) = 1 + \frac{1}{3} \gamma_{a0} \frac{\alpha_s}{4\pi}
+ \mathcal{O}(\alpha_s^2)\,.
\label{R1}
\end{equation}
The general result for $R(\alpha_s)$ can be derived by solving~(\ref{RGR})
and requiring the absence of fractional powers of $\alpha_s$
(or by requiring that~(\ref{R1}) is reproduced):
\begin{equation}
R(\alpha_s) = K_{\gamma_a}(\alpha_s) \left[ 1 + \alpha_s^{\frac{\gamma_{a0}}{2\beta_0}}
\int_0^{\alpha_s} \left( \frac{3\gamma_b(\alpha_s)}{2\beta(\alpha_s)} K_{-\gamma_a}(\alpha_s)
+ \frac{\gamma_{a0}}{2\beta_0} \right) \alpha_s^{-\frac{\gamma_{a0}}{2\beta_0}}
\frac{d\alpha_s}{\alpha_s} \right]\,.
\label{R2}
\end{equation}

The matrix element of $O^\alpha_5$ is~\cite{Ne:92}
\begin{equation}
{<}0|O^\alpha_5(\mu)|B^*{>} = \sqrt{m_{B^*}} F(\mu) G_k(\mu) e^\alpha\,.
\label{meO5}
\end{equation}
However, the formulae~\cite{Ne:92} for the matrix element of $O^\alpha_6$
hold only at the leading order.
Following~\cite{Ne:92}, we define
\begin{equation}
{<}0|O^m_3(\mu)|M{>} = \frac{1}{12} F(\mu) G(\mu) \Tr \Gamma
\frac{1+\rlap/v}{2} \sigma_{\mu\nu} \mathcal{M} \sigma^{\mu\nu}\,,
\label{meOm3}
\end{equation}
where $O^m_3$ is defined in~(\ref{Om}).
The $B$-meson matrix element $G_m(\mu)$~(\ref{Hmat1}) is,
at the next-to-leading order~(\ref{DOm3a}),
\begin{equation}
G_m(\mu) = G(\mu) +
\frac{1}{2} \left(\gamma^m_{10}+5\gamma^m_{20}\right)
\frac{\alpha_s(\mu)}{4\pi} \bar{\Lambda}\,.
\label{Gm}
\end{equation}
For $B^*$ we define
\begin{equation}
{<}0|O^\alpha_6|B^*{>} = - \frac{1}{3}
\left[ G_m(\mu) + R_m(\alpha_s(\mu)) \bar{\Lambda} \right]
\sqrt{m_{B^*}} F(\mu) e^\alpha\,,
\label{GmBst}
\end{equation}
where at the next-to-leading order~(\ref{DOm3b})
\begin{equation}
R_m(\alpha_s) = - \left(\gamma^m_{10}+4\gamma^m_{20}\right) \frac{\alpha_s}{4\pi}
+ \mathcal{O}(\alpha_s^2)\,.
\label{Rm}
\end{equation}
It obeys the renormalization-group equation
\begin{equation}
\frac{d\,R_m}{d\log\mu} + \gamma_m R_m + \gamma^m + \gamma^m_1 R + 3 \gamma^m_2 = 0\,.
\label{RGRm}
\end{equation}
Its solution (which contains no fractional powers of $\alpha_s$
and reproduces~(\ref{Rm})) is
\begin{eqnarray}
&&R_m(\alpha_s) = K_{\gamma_m}(\alpha_s) \alpha_s^{\frac{\gamma_{m0}}{2\beta_0}}
\label{Rm2}\\
&&\quad{}\times\int_0^{\alpha_s}
\frac{\gamma^m(\alpha_s) + \gamma^m_1(\alpha_s) R(\alpha_s) + 3 \gamma^m_2(\alpha_s)}
{2\beta(\alpha_s)}
K_{-\gamma_m}(\alpha_s) \alpha_s^{-\frac{\gamma_{m0}}{2\beta_0}}
\frac{d\alpha_s}{\alpha_s}\,.
\nonumber
\end{eqnarray}

Taking the matrix element of~(\ref{Exp}), we obtain
\begin{eqnarray}
&&\left\{\begin{array}{c}f_{B^*}\\f^T_{B^*}(\mu')\end{array}\right\} =
\frac{C_\Gamma(\mu',\mu) F(\mu)}{\sqrt{m_{B^*}}}
\nonumber\\
&&\qquad{} \times \left[ 1
+ \frac{1}{2m} \left( C^\Gamma_\Lambda(\mu) \bar{\Lambda}
+ G_k(\mu) - \tfrac{1}{3} C_m(\mu) G_m(\mu) \right) \right]\,,
\label{fBst1}
\end{eqnarray}
where $\Gamma=\gamma_\bot$, $\gamma_\bot\rlap/v$, and
\begin{eqnarray*}
&&C^\Gamma_\Lambda(\mu) = \pm 1 - \frac{B^\Gamma_4(\mu',\mu)}{C_\Gamma(\mu',\mu)}\\
&&\quad{} - \frac{1}{3} \left( 2
\left\{\begin{array}{c}C_\V(\mu',\mu)/C_{\gamma_\bot}(\mu',\mu)\\-1\end{array}\right\}
+ \frac{B^\Gamma_3(\mu',\mu)}{C_\Gamma(\mu',\mu)} \right) R(\alpha_s(\mu))\\
&&\quad{} - \frac{1}{3} C_m(\mu) R_m(\mu)\,.
\nonumber
\end{eqnarray*}
Substituting the solutions of the renormalization-group equations,
we arrive at the $\mu$-independent expressions
\begin{eqnarray}
&&\left\{\begin{array}{c}f_{B^*}\\\hat{f}^T_{B^*}\end{array}\right\} =
\left(\frac{\alpha_s(\mu_0)}{4\pi}\right)^{\frac{\tilde{\gamma}_0}{2\beta_0}}
\frac{\hat{C}_\Gamma \hat{F}}{\sqrt{m_{B^*}}}
\nonumber\\
&&\qquad{} \times \left[ 1
+ \frac{1}{2m} \left( \hat{C}^\Gamma_\Lambda \bar{\Lambda}
+ \hat{G}_k - \tfrac{1}{3} \hat{C}_m \hat{G}_m
\left(\frac{\alpha_s(\mu_0)}{4\pi}\right)^{\frac{\gamma_{m0}}{2\beta_0}}
\right) \right]\,,
\label{fBst2}
\end{eqnarray}
where
\begin{eqnarray}
&&\hat{C}^\Gamma_\Lambda = \pm 1 - \frac{B^\Gamma_4(\mu',\mu)}{C_\Gamma(\mu',\mu)}
\nonumber\\
&&\quad{} - \frac{1}{3} \left( 2
\left\{\begin{array}{c}C_\V(\mu',\mu)/C_{\gamma_\bot}(\mu',\mu)\\-1\end{array}\right\}
+ \frac{B^\Gamma_3(\mu',\mu)}{C_\Gamma(\mu',\mu)} \right) R(\alpha_s(\mu))
\nonumber\\
&&\quad{} - \frac{\gamma^k_0}{2\beta_0} \log \frac{\alpha_s(\mu)}{4\pi}
- \int_0^{\alpha_s(\mu)}
\left( \frac{\gamma^k(\alpha_s)}{2\beta(\alpha_s)}
- \frac{\gamma^k_0}{2\beta_0} \right)
\frac{d\alpha_s}{\alpha_s}
- \frac{1}{3} C_m(\mu) R_m(\alpha_s(\mu))
\nonumber\\
&&\quad{} + \frac{1}{3} \hat{C}_m \alpha_s(\mu_0)^{\frac{\gamma_{m0}}{2\beta_0}}
\int_0^{\alpha_s(\mu)} \frac{\gamma^m(\alpha_s)}{2\beta(\alpha_s)}
K_{-\gamma_m}(\alpha_s) \alpha_s^{-\frac{\gamma_{m0}}{2\beta_0}}
\frac{d\alpha_s}{\alpha_s}\,.
\label{hatC1}
\end{eqnarray}
At first sight, it is not obvious that $\hat{C}^\Gamma_\Lambda$
does not depend on $\mu$.
However, differentiating it in $\log\mu$ and taking into account
the renormalization-group equations~(\ref{RGB34}), (\ref{RGR}), (\ref{RGRm}),
we obtain zero.
Therefore, the expression with $\mu=\mu_0$ can be used:
\begin{eqnarray}
&&\hat{C}^\Gamma_\Lambda = \pm 1 - \frac{\hat{B}^\Gamma_4}{\hat{C}_\Gamma}
- \frac{1}{3} \left( 2
\left\{\begin{array}{c}\hat{C}_\V/\hat{C}_{\gamma_\bot}\\-1\end{array}\right\}
+ \frac{\hat{B}^\Gamma_3}{\hat{C}_\Gamma} K_{-\gamma_a}(\alpha_s(\mu_0)) \right)
R(\alpha_s(\mu_0))
\nonumber\\
&&\quad{} - \frac{\gamma^k_0}{2\beta_0} \log \frac{\alpha_s(\mu_0)}{4\pi}
- \int_0^{\alpha_s(\mu_0)}
\left( \frac{\gamma^k(\alpha_s)}{2\beta(\alpha_s)}
- \frac{\gamma^k_0}{2\beta_0} \right)
\frac{d\alpha_s}{\alpha_s}
\nonumber\\
&&\quad{} - \frac{1}{3} \hat{C}_m K_{-\gamma_m}(\alpha_s(\mu_0)) R_m(\alpha_s(\mu_0))
\nonumber\\
&&\quad{} + \frac{1}{3} \hat{C}_m \alpha_s(\mu_0)^{\frac{\gamma_{m0}}{2\beta_0}}
\int_0^{\alpha_s(\mu_0)} \frac{\gamma^m(\alpha_s)}{2\beta(\alpha_s)}
K_{-\gamma_m}(\alpha_s) \alpha_s^{-\frac{\gamma_{m0}}{2\beta_0}}
\frac{d\alpha_s}{\alpha_s}\,.
\label{hatC2}
\end{eqnarray}
Comparing this with~(\ref{hatC1}), we can rewrite $B^\Gamma_4(\mu',\mu)/C_\Gamma(\mu',\mu)$
in a form which seems different from~(\ref{B4}) but is equal to it:
\begin{eqnarray}
&&\frac{B^\Gamma_4(\mu',\mu)}{C_\Gamma(\mu',\mu)} =
\frac{\hat{B}^\Gamma_4}{\hat{C}_\Gamma}
- \frac{1}{3} \left( 2
\left\{\begin{array}{c}\hat{C}_\V/\hat{C}_{\gamma_\bot}\\-1\end{array}\right\}
+ \frac{B^\Gamma_3(\mu',\mu)}{C_\Gamma(\mu',\mu)} \right) R(\alpha_s(\mu))
\nonumber\\
&&\quad{} + \frac{1}{3} \left( 2
\left\{\begin{array}{c}\hat{C}_\V/\hat{C}_{\gamma_\bot}\\-1\end{array}\right\}
+ \frac{\hat{B}^\Gamma_3}{\hat{C}_\Gamma} K_{-\gamma_a}(\alpha_s(\mu_0)) \right)
R(\alpha_s(\mu_0))
\nonumber\\
&&\quad{} - \frac{\gamma^k_0}{2\beta_0}
\log \frac{\alpha_s(\mu)}{\alpha_s(\mu_0)}
- \int_{\alpha_s(\mu_0)}^{\alpha_s(\mu)}
\left( \frac{\gamma^k(\alpha_s)}{2\beta(\alpha_s)}
- \frac{\gamma^k_0}{2\beta_0} \right)
\frac{d\alpha_s}{\alpha_s}
\nonumber\\
&&\quad{} - \frac{1}{3} C_m(\mu) R_m(\alpha_s(\mu))
+ \frac{1}{3} \hat{C}_m K_{-\gamma_m}(\alpha_s(\mu_0)) R_m(\alpha_s(\mu_0))
\nonumber\\
&&\quad{} + \frac{1}{3} \hat{C}_m \alpha_s(\mu_0)^{\frac{\gamma_{m0}}{2\beta_0}}
\int_{\alpha_s(\mu_0)}^{\alpha_s(\mu)} \frac{\gamma^m(\alpha_s)}{2\beta(\alpha_s)}
K_{-\gamma_m}(\alpha_s) \alpha_s^{-\frac{\gamma_{m0}}{2\beta_0}}
\frac{d\alpha_s}{\alpha_s}
\label{B4a}
\end{eqnarray}
(to convince oneself that this is equivalent to~(\ref{B4}),
one can check that they coincide at $\mu=\mu_0$,
and that~(\ref{B4a}) obeys the renormalization-group equation~(\ref{RGB34}).)

At the next-to-leading order,
\begin{eqnarray}
&&\left\{\begin{array}{c}f_{B^*}\\\hat{f}^T_{B^*}\end{array}\right\} =
\left(\frac{\alpha_s(\mu_0)}{4\pi}\right)^{\frac{\tilde{\gamma}_0}{2\beta_0}}
\frac{\hat{C}_\Gamma \hat{F}}{\sqrt{m_{B^*}}}
\Biggl\{ 1
\nonumber\\
&&\quad{} + \frac{1}{2m} \Biggl[
\left( - \frac{\gamma^k_0}{2\beta_0} \log \frac{\alpha_s(\mu_0)}{4\pi}
+ \frac{1}{3} \left( \pm 1 - \frac{\gamma^m_0}{\gamma_{m0}} \right)
+ c^\Gamma_{\Lambda1} \frac{\alpha_s(\mu_0)}{4\pi} + \cdots \right) \bar{\Lambda}
\nonumber\\
&&\qquad{} + \hat{G}_k - \frac{1}{3} \hat{G}_m
\left(\frac{\alpha_s(\mu_0)}{4\pi}\right)^{\frac{\gamma_{m0}}{2\beta_0}}
\left( 1 + c_{m1} \frac{\alpha_s(\mu_0)}{4\pi} + \cdots \right)
\Biggr] \Biggr\}\,,
\label{fBst3}
\end{eqnarray}
where
\begin{eqnarray*}
&&c^\Gamma_{\Lambda1} = \tfrac{1}{3} (1\pm1) \left(c^{\gamma_\bot}_1-c^\V_1\right)
- \tfrac{1}{3} b^\Gamma_{31} - b^\Gamma_{41}
- \tfrac{1}{3} \frac{\gamma^m_0}{\gamma_{m0}} c_{m1}
\mp \tfrac{2}{9} \gamma_{a0}
+ \tfrac{1}{3} \gamma^m_{10} + \tfrac{4}{3} \gamma^m_{20}\\
&&\quad{} - \frac{\gamma^k_1-\frac{1}{3}\gamma^m_1}{2\beta_0}
+ \frac{\beta_1\left(\gamma^k_0-\frac{1}{3}\gamma^m_0\right)}
{2\beta_0^2}
- \frac{\gamma_{m0}\gamma^m_1-\gamma^m_0\gamma_{m1}}
{6\beta_0\left(\gamma_{m0}-2\beta_0\right)}\,.
\end{eqnarray*}
In the large-$\beta_0$ limit,
the IR renormalon ambiguities $\Delta \hat{C}_\Gamma$~(\ref{DCLb0})
for $\Gamma=\gamma_\bot$, $\gamma_\bot\rlap/v$
(having $n=1$, $\eta=-1$ and $n=2$, $\eta=1$)
are $(11/12)(\Delta\bar{\Lambda}/m)$ and $(5/4)(\Delta\bar{\Lambda}/m)$.
They cancel with the UV renormalon ambiguities
of the hadronic matrix elements~(\ref{DLbar}), (\ref{DGhat})
in the $1/m$ correction in~(\ref{fBst3}).

The Borel images are
\begin{eqnarray}
&&S_\Gamma(u) =
\frac{C_F}{\left(\frac{1}{2}-u\right)^{1+\frac{\beta_1}{2\beta_0^2}}}
\left[1+c^{\Gamma\prime}_1\left(\tfrac{1}{2}-u\right)+\cdots\right]
\Biggl\{
\nonumber\\
&&\quad\Biggl[ - \frac{\gamma^k_0}{2\beta_0}
\left( \log \frac{\frac{1}{2}-u}{\beta_0}
- \psi\left(1+\frac{\beta_1}{2\beta_0^2}\right) \right)
\nonumber\\
&&\qquad{} + \frac{1}{3} \left(\pm1-\frac{\gamma^m_0}{\gamma_{m0}}\right)
+ c^{\Gamma\prime}_{\Lambda1} \left(\tfrac{1}{2}-u\right) + \cdots
\Biggr] N_0'
\nonumber\\
&&\quad{} - \frac{3}{2} N_1'
- \frac{1}{3} \left(2-\frac{\gamma^m_0}{\gamma_{m0}}\right) N_2'
\left(\tfrac{1}{2}-u\right)^{\frac{\gamma_{m0}}{2\beta_0}}
\left[1 + c_{m1}'\left(\tfrac{1}{2}-u\right) + \cdots\right]
\Biggr\}\,,
\label{Su2}
\end{eqnarray}
(see~(\ref{Npr})), and the perturbative coefficients at $n\gg1$ are
\begin{eqnarray}
&&c^\Gamma_{n+1} = 2 C_F\,n!\,(2\beta_0)^n\,(2 \beta_0 n)^{\frac{\beta_1}{2\beta_0^2}}
\left(1 + \frac{c^{\Gamma\prime\prime}_1}{2 \beta_0 n} + \cdots\right)
\Biggl[
\nonumber\\
&&\quad\left( \frac{\gamma^k_0}{2\beta_0} \log 2 \beta_0 n
+ \frac{1}{3} \left(\pm1-\frac{\gamma^m_0}{\gamma_{m0}}\right)
+ \frac{c^{\Gamma\prime\prime}_{\Lambda1}}{2 \beta_0 n}
+ \cdots \right) N_0
\nonumber\\
&&\qquad{} - \frac{3}{2} N_1
- \frac{1}{3} \left(2-\frac{\gamma^m_0}{\gamma_{m0}}\right) N_2
(2 \beta_0 n)^{-\frac{\gamma_{m0}}{2\beta_0}}
\left(1 + \frac{c_{m1}''}{2 \beta_0 n} + \cdots
\right) \Biggr]
\label{asy2}
\end{eqnarray}
(see~(\ref{Nprpr})).

The ratio $\hat{f}^T_{B^*}/f_{B^*}$ at the leading order in $1/m$
is given by the perturbative series in $\alpha_s(\mu_0)$ with
\begin{eqnarray}
&&c^{(\hat{f}^T_{B^*}/f_{B^*})}_1 = C_F \left[ - \tfrac{1}{2}
+ \left(\tfrac{25}{2}C_F-\tfrac{4}{3}C_A\right) \frac{1}{\beta_0'}
- \left(11C_F+7C_A\right) \frac{C_A}{\beta_0^{\prime2}} \right]\,,
\label{cst}\\
&&c^{(\hat{f}^T_{B^*}/f_{B^*})}_2 = C_F \Biggl\{
- \left(\tfrac{1}{3}\pi^2+\tfrac{1}{4}\right) \beta_0'
+ \left(2\zeta_3+\tfrac{8}{3}\pi^2\log2-\tfrac{35}{9}\pi^2+\tfrac{139}{24}\right) C_F
\nonumber\\
&&\quad{}
+ \left(-4\zeta_3-\tfrac{4}{3}\pi^2\log2+\tfrac{23}{9}\pi^2
-\tfrac{135}{16}\right) C_A
+ \left(-\tfrac{4}{3}\pi^2+\tfrac{46}{9}\right) T_F
\nonumber\\
&&\quad{} + \biggl[ \left(32\zeta_3-\tfrac{337}{6}\right) C_F^2
+ \left(-78\zeta_3+\tfrac{151}{4}\right) C_F C_A
+ \left(42\zeta_3-\tfrac{131}{36}\right) C_A^2
\nonumber\\
&&\qquad{} - \tfrac{250}{9} C_F T_F + \tfrac{80}{27} C_A T_F
\biggr] \frac{1}{\beta_0'}
\nonumber\\
&&\quad{} + \Bigl( \tfrac{625}{8} C_F^3 + \tfrac{461}{6} C_F^2 C_A
+ \tfrac{2375}{36} C_F C_A^2 - \tfrac{511}{48} C_A^3
\nonumber\\
&&\qquad{} + \tfrac{220}{9} C_F C_A T_F + \tfrac{140}{9} C_A^2 T_F \Bigr)
\frac{1}{\beta_0^{\prime2}}
\nonumber\\
&&\quad{} - \tfrac{1}{6} \left( 75 C_F^2 + 25 C_F C_A + 21 C_A^2 \right)
\left( 11 C_F + 7 C_A \right) \frac{C_A}{\beta_0^{\prime3}}
\nonumber\\
&&\quad{} + \tfrac{1}{2} C_F \left( 11 C_F + 7 C_A \right)^2 \frac{C_A^2}{\beta_0^{\prime4}}
\Biggr\}
\nonumber
\end{eqnarray}
(from the result in~\cite{BG:95}, omitting the $m_c\ne0$ effect,
and the three-loop anomalous dimension $\gamma'_\sigma$ of the tensor current~\cite{Gr:00}).
This ratio is, from~(\ref{fBst2}),
\begin{eqnarray}
&&\frac{\hat{f}^T_{B^*}}{f_{B^*}} =
\frac{\hat{C}_{\gamma_\bot\V}}{\hat{C}_{\gamma_\bot}}
\left[1 - \frac{\bar{\Lambda}}{3m}
\left( 1 + c^{(\hat{f}^T_{B^*}/f_{B^*})}_{\Lambda1} \frac{\alpha_s(\mu_0)}{4\pi}
+ \cdots \right) \right]\,,
\label{rst}\\
&&c^{(\hat{f}^T_{B^*}/f_{B^*})}_{\Lambda1} =
\tfrac{3}{2} \left( c^{\gamma_\bot\V}_{\Lambda1} - c^{\gamma_\bot}_{\Lambda1} \right)
%= c^{\gamma_\bot}_1 - c^\V_1
%+ \tfrac{1}{2} \left(b^{\gamma_\bot\V}_{31}-b^{\gamma_\bot}_{31}\right)
%+ \tfrac{3}{2} \left(b^{\gamma_\bot\V}_{41}-b^{\gamma_\bot}_{41}\right)
%- \tfrac{2}{3} \gamma_{a0}\,.
= - 8 C_F\,.
\nonumber
\end{eqnarray}
Therefore, the Borel image of the perturbative series is
\begin{eqnarray}
&&S_{\hat{f}^T_{B^*}/f_{B^*}}(u) =
- \frac{2}{3} \frac{C_F N_0'}{\left(\frac{1}{2}-u\right)^{1+\frac{\beta_1}{2\beta_0^2}}}
\left[1 + c^{(\hat{f}^T_{B^*}/f_{B^*})\prime}_1\left(\tfrac{1}{2}-u\right) + \cdots\right]\,,
\label{Sust}\\
&&c^{(\hat{f}^T_{B^*}/f_{B^*})\prime}_1 = \frac{2\beta_0}{\beta_1}
\left(c^{\hat{f}^T_{B^*}/f_{B^*}}_{\Lambda1} + c^{\hat{f}^T_{B^*}/f_{B^*}}_1
- \frac{\beta_0\beta_2-\beta_1^2}{2\beta_0^3}\right)\,,
\nonumber
\end{eqnarray}
and the asymptotics of $c^{(\hat{f}^T_{B^*}/f_{B^*})}_L$ at $L\gg1$ is
\begin{eqnarray}
&&c^{(\hat{f}^T_{B^*}/f_{B^*})}_{n+1} =
- \frac{4}{3} C_F N_0\,n!\,(2\beta_0)^n\,(2 \beta_0 n)^{\frac{\beta_1}{2\beta_0^2}}
\left(1 + \frac{c^{(\hat{f}^T_{B^*}/f_{B^*})\prime\prime}_1}{2\beta_0 n} + \cdots\right)\,,
\label{cnst}\\
&&c^{(\hat{f}^T_{B^*}/f_{B^*})\prime\prime}_1 =
c^{(\hat{f}^T_{B^*}/f_{B^*})}_{\Lambda1} + c^{(\hat{f}^T_{B^*}/f_{B^*})}_1
- \frac{2\beta_0\beta_2-2\beta_0^2\beta_1-3\beta_1^2}{4\beta_0^3}\,.
\nonumber
\end{eqnarray}

\section{Results and conclusion}
\label{Conc}

Our main result is for the ratio of two (in principle) measurable quantities,
$f_{B^*}/f_B$.
It is given by the perturbative series in $\alpha_s(\mu_0)$ with
\begin{eqnarray}
&&c^{(f_{B^*}/f_B)}_1 = - 2 C_F\,,
\label{cff}\\
&&c^{(f_{B^*}/f_B)}_2 = C_F \biggl[ - 3 \beta'_0
+ \left( - 8\zeta_3 + \tfrac{16}{3}\pi^2\log2 - \tfrac{64}{9}\pi^2 + \tfrac{31}{3} \right) C_F
\nonumber\\
&&\quad{} + \left( 4\zeta_3 -\tfrac{8}{3}\pi^2\log2 + \tfrac{8}{3}\pi^2 - 6 \right) C_A
+ \left( \tfrac{16}{9} \pi^2 - \tfrac{200}{9} \right) T_F \biggr]\,.
\nonumber
\end{eqnarray}
(from~\cite{BG:95}, omitting the $m_c\ne0$ effect).
This ratio is, from~(\ref{fB3}) and~(\ref{fBst3}),
\begin{eqnarray}
&&\frac{f_{B^*}}{f_B} =
\frac{\hat{C}_{\gamma_\bot}}{\hat{C}_\V}
\Biggl\{ 1 + \frac{2}{3m} \Biggl[
\left( 1 - \frac{\gamma^m_0}{\gamma_{m0}}
+ c^{(f_{B^*}/f_B)}_{\Lambda1} \frac{\alpha_s(\mu_0)}{4\pi} + \cdots \right) \bar{\Lambda}
\nonumber\\
&&\quad{} - \hat{G}_m
\left(\frac{\alpha_s(\mu_0)}{4\pi}\right)^{\frac{\gamma_{m0}}{2\beta_0}}
\left( 1 + c_{m1} \frac{\alpha_s(\mu_0)}{4\pi} + \cdots \right)
\Biggr] \Biggr\}\,,
\label{rff}
\end{eqnarray}
where
\begin{eqnarray*}
&&c^{(f_{B^*}/f_B)}_{\Lambda1}
= \frac{3}{4} \left( c^{\gamma_\bot}_{\Lambda1} - c^\V_{\Lambda1} \right)\\
&&\quad{} = 6 C_F \left[ - 1 - \frac{4}{3} \frac{C_F}{C_A}
+ \frac{8(4\pi^2+3)C_F-(8\pi^2-93)C_A}{27(\beta_0-C_A)} \right]\,.
\end{eqnarray*}
The Borel image of the perturbative series is
\begin{eqnarray}
&&S_{f_{B^*}/f_B}(u) = \frac{4}{3}\,
\frac{C_F}{\left(\frac{1}{2}-u\right)^{1+\frac{\beta_1}{2\beta_0^2}}}
\Biggl[
\left(1 - \frac{\gamma^m_0}{\gamma_{m0}}
+ c^{(f_{B^*}/f_B)\prime}_{\Lambda1} \left(\tfrac{1}{2}-u\right) + \cdots \right) N_0'
\nonumber\\
&&\quad{} - \left(2-\frac{\gamma^m_0}{\gamma_{m0}}\right)
N_2' \left(\tfrac{1}{2}-u\right)^{\frac{\gamma_{m0}}{2\beta_0}}
\left(1 + c^{(f_{B^*}/f_B)\prime}_{m1} \left(\tfrac{1}{2}-u\right) + \cdots \right)
\Biggr]\,,
\label{Suff}
\end{eqnarray}
where
\begin{eqnarray*}
&&c^{(f_{B^*}/f_B)\prime}_{\Lambda1} = \frac{2\beta_0}{\beta_1}
\left[ c^{(f_{B^*}/f_B)}_{\Lambda1}
+ \left(1-\frac{\gamma^m_0}{\gamma_{m0}}\right) c^{(f_{B^*}/f_B)\prime}_1
\right]\,,\\
&&c^{(f_{B^*}/f_B)\prime}_{m1} = \frac{2\beta_0}{\beta_1-\beta_0\gamma_{m0}}
\left( c_{m1} + c^{(f_{B^*}/f_B)\prime}_1 \right)\,,\\
&&c^{(f_{B^*}/f_B)\prime}_1 = c^{(f_{B^*}/f_B)}_1
- \frac{\beta_0\beta_2-\beta_1^2}{2\beta_0^3}\,.
\end{eqnarray*}
The asymptotics of the coefficients is
\begin{eqnarray}
&&c^{(f_{B^*}/f_B)}_{n+1} = \frac{8}{3} C_F\,n!\,(2\beta_0)^n
(2 \beta_0 n)^{\frac{\beta_1}{2\beta_0^2}} \Biggl[
\left( 1 - \frac{\gamma^m_0}{\gamma_{m0}}
+ \frac{c^{(f_{B^*}/f_B)\prime\prime}_{\Lambda1}}{2\beta_0 n} + \cdots \right) N_0
\nonumber\\
&&\quad{} - \left(2-\frac{\gamma^m_0}{\gamma_{m0}}\right) N_2
(2 \beta_0 n)^{-\frac{\gamma_{m0}}{2\beta_0}}
\left(1 + \frac{c^{(f_{B^*}/f_B)\prime\prime}_{m1}}{2\beta_0 n} + \cdots\right)
\Biggr]\,,
\label{cnff}
\end{eqnarray}
where
\begin{eqnarray*}
&&c^{(f_{B^*}/f_B)\prime\prime}_{\Lambda1} = c^{(f_{B^*}/f_B)}_{\Lambda1}
+ \left(1-\frac{\gamma^m_0}{\gamma_{m0}}\right)
\left( c^{(f_{B^*}/f_B)\prime}_1 + \frac{\beta_1(\beta_1+2\beta_0^2)}{4\beta_0^3} \right)\,,\\
&&c^{(f_{B^*}/f_B)\prime\prime}_{m1} = c_{m1} + c^{(f_{B^*}/f_B)\prime}_1 +
\frac{(\beta_1-\beta_0\gamma_{m0})(\beta_1+2\beta_0^2-\beta_0\gamma_{m0})}{4\beta_0^3}\,.
\end{eqnarray*}

Substituting the parameters, we have the perturbative series
\begin{eqnarray}
&&\frac{f_{B^*}}{f_B} = 1 - \frac{2}{3} \frac{\alpha_s(\mu_0)}{\pi}
- \left( - \frac{1}{9} \zeta_3 + \frac{2}{27} \pi^2 \log 2
+ \frac{4}{81} \pi^2 + \frac{115}{36} \right)
\left(\frac{\alpha_s(\mu_0)}{\pi}\right)^2
\nonumber\\
&&\quad{} + \cdots
+ \mathcal{O}\left(\frac{\Lambda_{\overline{\mathrm{MS}}}}{m_b}\right)\,.
\label{fExN}
\end{eqnarray}
The asymptotics~(\ref{cnff}) becomes
\begin{eqnarray}
&&c^{(f_{B^*}/f_B)}_{n+1} = - \frac{224}{81}\,n!\,
\left(\frac{50}{3}\right)^n \left(\frac{50}{3}n\right)^{\frac{231}{625}}
\Biggl\{ \left[ 1 - \frac{2}{25} \left(\pi^2+\frac{924493}{250000}\right) \frac{1}{n}
+ \cdots \right] N_0
\nonumber\\
&&\quad{} + \frac{2}{7} N_2 \left(\frac{50}{3}n\right)^{-\frac{9}{25}}
\left(1 + \frac{40157}{3125000} \frac{1}{n} + \cdots \right) \Biggr\}\,.
\label{fAsN}
\end{eqnarray}
Definite quantitative predictions cannot be made,
because the normalization coefficients $N_{0,2}$ are unknown.
To have some idea about the growth of the coefficients,
we present them in Table~\ref{T:f}.
The coefficients $c_L/4^L$ of $\alpha_s(\mu_0)/\pi$ are given in three columns.
The first column shows the exactly known ones~(\ref{fExN}).
The second column shows the results of the large-$\beta_0$ limit.
The Borel image $S_{f_{B^*}/f_B}(u)$ in this limit is, from~(\ref{Lb0}),
\begin{equation}
S_{f_{B^*}/f_B}(u) = - 4 C_F \frac{\Gamma(1+u)\Gamma(1-2u)}{\Gamma(3-u)}\,.
\label{fLb0}
\end{equation}
Expanding it at $u=0$~(\ref{Borel}), we get
\begin{eqnarray*}
&&c^{(f_{B^*}/f_B)}_1 = - 2 C_F\,,\quad
c^{(f_{B^*}/f_B)}_2 = - 3 C_F \beta_0\,,\quad
c^{(f_{B^*}/f_B)}_3 = - C_F \left( \tfrac{4}{3} \pi^2 + 7 \right) \beta_0^2\,,\\
&&c^{(f_{B^*}/f_B)}_4 = - 3 C_F \left( 8 \zeta_3 + 2 \pi^2 + \tfrac{15}{2} \right) \beta_0^3\,,\\
&&c^{(f_{B^*}/f_B)}_5 = - C_F \left( 144 \zeta_3 + \tfrac{24}{5} \pi^4 + 28 \pi^2 + 93
\right) \beta_0^4\,,\\
&&c^{(f_{B^*}/f_B)}_6 = - C_F \left( 1440 \zeta_5 + 160 \pi^2 \zeta_3 + 840 \zeta_3
+ 36 \pi^4 + 150 \pi^2 + \tfrac{945}{2} \right) \beta_0^5\,,
\ldots
\end{eqnarray*}
This limit reproduces $c^{(f_{B^*}/f_B)}_1$
and the $\beta_0$-term of $c^{(f_{B^*}/f_B)}_2$~(\ref{cff}).
Finally, the third column shows the asymptotics~(\ref{fAsN})
at $N_0=N_2=1$.
Let's stress once more that this is \emph{not} the result of QCD,
but simply a numerical illustration of the typical behaviour
of the perturbative coefficients.
The large-$\beta_0$ result includes not just the pole at $u=\frac{1}{2}$,
but the whole function $S_{f_{B^*}/f_B}(u)$~(\ref{fLb0}).
In contrast to this, the asymptotics~(\ref{fAsN})
is determined by the nearest singularity at $u=\frac{1}{2}$ only,
but includes all powers of $\beta_0$, not just the highest one.
To show the rate of convergence of the $1/n$ expansion~(\ref{fAsN}),
the asymptotic result is also expressed via the leading term
and the $1/n$ correction.
One can see that the accuracy of our next-to-leading order results
at $L\lesssim10$ is not high.
Finally, the last column shows the complete $L$-loop contribution
to $f_{B^*}/f_B$, according to~(\ref{fAsN}) with $N_0=N_2=1$.
We use the value $\alpha_s(e^{-5/6}m_b)=0.299$
obtained by RunDec~\cite{CKS:00}.
The smallest contribution seems to be the 3-loop one,
and it is about $4\%$
(though this small value is due to a partial cancellation
between the leading order and the next-to-leading one,
and thus is not quite reliable).
Therefore, calculation of this 3-loop correction is meaningful
(and it is actually possible, using the technique of~\cite{MR:00,MR:00b}),
while there would be no sense in the 4-loop calculation.

\begin{table}[ht]
\caption{The perturbative series for $f_{B^*}/f_B$}
\label{T:f}
\vspace{1mm}
\begin{tabular}{|c|r@{.}l|r@{.}l|r@{.}l@{${}={}$}r@{${}\cdot(1-0.$}l@{$)$\hspace{0.5\columnsep}}|r@{.}l|}
\hline
\raisebox{-4mm}[0mm][0mm]{$L$} & \multicolumn{8}{c|}{$-c^{(f_{B^*}/f_B)}_L/4^L$}
& \multicolumn{2}{c|}{\raisebox{-4mm}[0mm][0mm]{$-{}$cont.}} \\
\cline{2-9}
& \multicolumn{2}{c|}{exact} & \multicolumn{2}{c|}{large $\beta_0$} &
\multicolumn{4}{c|}{asymptotic} &\multicolumn{2}{c|}{} \\
\hline
1 & 0&67 & 0&67 & \multicolumn{4}{c|}{} & \multicolumn{2}{c|}{} \\
2 & 4&06 & 2&08 & \multicolumn{4}{c|}{} & \multicolumn{2}{c|}{} \\
3 & \multicolumn{2}{c|}{} &     29&17 &     47&26&     95&501 & 0&0406 \\
4 & \multicolumn{2}{c|}{} &    333&26 &    902&44&   1363&338 & 0&0736 \\
5 & \multicolumn{2}{c|}{} &   6342&19 &  18699&42&  25103&255 & 0&1450 \\
6 & \multicolumn{2}{c|}{} & 128998&30 & 449431&53& 565332&205 & 0&3311 \\
\hline
\end{tabular}
\end{table}

For the ratio $m/\hat{m}$, we obtain (from the Appendix~\ref{MM})
\begin{eqnarray}
&&\frac{m}{\hat{m}} = 1
+ \tfrac{891}{1058} \frac{\alpha_s(\mu_0)}{\pi}
+ \left( - \tfrac{173}{138} \zeta_3 + \tfrac{1}{9} \pi^2 \log 2 + \tfrac{1}{9} \pi^2
+ \tfrac{168550145}{40297104} \right)
\left(\frac{\alpha_s(\mu_0)}{\pi}\right)^2
\nonumber\\
&&\quad{} + \Bigl( - \tfrac{188}{27} a_4 + \tfrac{50225}{4968} \zeta_5
- \tfrac{1439}{432} \pi^2 \zeta_3 - \tfrac{4396763}{438012} \zeta_3
- \tfrac{47}{162} \log^4 2 - \tfrac{14}{81} \pi^2 \log^2 2
\nonumber\\
&&\qquad{} - \tfrac{402485}{85698} \pi^2 \log 2 + \tfrac{461}{7776} \pi^4
+ \tfrac{220317449}{20567520} \pi^2 + \tfrac{17341442069927}{767418048576}
\Bigr)
\left(\frac{\alpha_s(\mu_0)}{\pi}\right)^3
\nonumber\\
&&\quad{} + \cdots
+ \mathcal{O}\left(\frac{\Lambda_{\overline{\mathrm{MS}}}}{m_b}\right)\,.
\label{mExN}
\end{eqnarray}
The asymptotics~(\ref{cnmm}) becomes
\begin{eqnarray}
&&c^{(m/\hat{m})}_{n+1} = \frac{16}{3} N_0\,n!\,
\left(\frac{50}{3}\right)^n \left(\frac{50}{3}n\right)^{\frac{231}{625}}
\Biggl[ 1 + \frac{688161953}{1653125000} \frac{1}{n}
\label{mAsN}\\
&&\quad{} + \left( - \frac{880261}{7187500} \zeta_3
+ \frac{4}{625} \pi^2 \log 2 + \frac{4}{625} \pi^2
+ \frac{8332134087653830381}{49190800781250000000} \right) \frac{1}{n^2}
\Biggr]\,.
\nonumber
\end{eqnarray}
It depends on just one normalization constant $N_0$.
In the large-$\beta_0$ limit, from~(\ref{Lb0}) we obtain~\cite{BB:94}
\begin{equation}
S_{m/\hat{m}}(u) = 6 C_F
\left[ \frac{\Gamma(u)\Gamma(1-2u)}{\Gamma(3-u)} (1-u) - \frac{1}{2u} \right]\,,
\label{mLb0}
\end{equation}
and
\begin{eqnarray*}
&&c^{(m/\hat{m})}_1 = \tfrac{3}{2} C_F\,,\quad
c^{(m/\hat{m})}_2 = C_F \left( \pi^2 + \tfrac{3}{4} \right) \beta_0\,,\quad
c^{(m/\hat{m})}_3 = C_F \left( 12 \zeta_3 + \pi^2 + \tfrac{3}{4} \right) \beta_0^2\,,\\
&&c^{(m/\hat{m})}_4 = 3 C_F \left( 6 \zeta_3 + \tfrac{3}{5} \pi^4 + \tfrac{1}{2} \pi^2
+ \tfrac{3}{8} \right) \beta_0^3\,,\\
&&c^{(m/\hat{m})}_5 = 3 C_F \left( 144 \zeta_5 + 16 \pi^2 \zeta_3 + 12 \zeta_3
+ \tfrac{6}{5} \pi^4 + \pi^2 + \tfrac{3}{4} \right) \beta_0^4\,,\\
&&c^{(m/\hat{m})}_6 = C_F \left( 720 \zeta_3^2 + 1080 \zeta_5 + 120 \pi^2 \zeta_3 + 90 \zeta_3
+ \tfrac{244}{21} \pi^6 + 9 \pi^4 + \tfrac{15}{2} \pi^2 + \tfrac{45}{8} \right) \beta_0^5\,,
\ldots
\end{eqnarray*}
(the terms with the highest powers of $\beta_0$ in the Appendix~\ref{MM} are reproduced).
Numerical results are shown in Table~\ref{T:m}.
For $L=3$, the $1/(L-1)$ expansion~(\ref{mAsN}) seems to converge well;
comparison with the exact 3-loop result from Appendix~\ref{MM} suggests
that the normalization factor $N_0$ is smaller that its large-$\beta_0$ value 1,
namely, $N_0\approx0.27$.
This conclusion is in a qualitative agreement with the estimate~(\ref{pineda}),
especially if we omit the problematic 3-loop correction in it.

\begin{table}[ht]
\caption{The perturbative series for $m/\hat{m}$}
\label{T:m}
\vspace{1mm}
\begin{tabular}{|c|r@{.}l|r@{.}l|r@{.}l@{${}={}$}r@{${}\cdot(1+0.$}l@{${}+0.0$}l
@{$)$\hspace{0.5\columnsep}}|r@{.}l|}
\hline
\raisebox{-4mm}[0mm][0mm]{$L$} & \multicolumn{9}{c|}{$c^{(m/\hat{m})}_L/4^L$}
& \multicolumn{2}{c|}{\raisebox{-4mm}[0mm][0mm]{cont.}} \\
\cline{2-10}
& \multicolumn{2}{c|}{exact} & \multicolumn{2}{c|}{large $\beta_0$} &
\multicolumn{5}{c|}{asymptotic} &\multicolumn{2}{c|}{} \\
\hline
1 &  0&84 &  0&5  & \multicolumn{5}{c|}{} & \multicolumn{2}{c|}{} \\
2 &  4&53 &  7&37 & \multicolumn{5}{c|}{} & \multicolumn{2}{c|}{} \\
3 & 56&37 & 36&23 & 209&88 & 169&208&32 & 0&180 \\
4 & \multicolumn{2}{c|}{} &    641&71 &    2833&12&    2457&139&14 & 0&231 \\
5 & \multicolumn{2}{c|}{} &   9062&48 &   50650&18&   45543&104&08 & 0&393 \\
6 & \multicolumn{2}{c|}{} & 206941&30 & 1121489&20& 1030382&083&05 & 0&826 \\
\hline
\end{tabular}
\end{table}

Finally, our last ratio~(\ref{cst}) is
\begin{eqnarray}
&&\frac{\hat{f}^T_{B^*}}{f_{B^*}} = 1
- \tfrac{707}{3174} \frac{\alpha_s(\mu_0)}{\pi}
- \left( - \tfrac{77}{138} \zeta_3 + \tfrac{1}{27} \pi^2 \log 2 + \tfrac{5}{81} \pi^2
+ \tfrac{102436609}{120891312} \right)
\left(\frac{\alpha_s(\mu_0)}{\pi}\right)^2
\nonumber\\
&&\quad{} + \cdots
+ \mathcal{O}\left(\frac{\Lambda_{\overline{\mathrm{MS}}}}{m_b}\right)\,.
\label{sExN}
\end{eqnarray}
The asymptotics~(\ref{cnst}) becomes
\begin{equation}
c^{(\hat{f}^T_{B^*}/f_{B^*})}_{n+1} = - \frac{16}{9} N_0\,n!\,
\left(\frac{50}{3}\right)^n \left(\frac{50}{3}n\right)^{\frac{231}{625}}
\left( 1 - \frac{792338047}{1653125000} \frac{1}{n} + \cdots \right)\,.
\label{sAsN}
\end{equation}
It also depends on just one normalization constant $N_0$.
In the large-$\beta_0$ limit, from~(\ref{Lb0}),
$S_{\hat{f}^T_{B^*}/f_{B^*}}(u)$ differs from~(\ref{mLb0}) by the factor $-\frac{1}{3}$
(this reproduces the terms with the highest powers of $\beta_0$ in~(\ref{cnst})).
Numerical results are shown in Table~\ref{T:s}.

\begin{table}[ht]
\caption{The perturbative series for $\hat{f}^T_{B^*}/f_{B^*}$}
\label{T:s}
\vspace{1mm}
\begin{tabular}{|c|r@{.}l|r@{.}l|r@{.}l@{${}={}$}r@{${}\cdot(1-0.$}l@{$)$\hspace{0.5\columnsep}}|r@{.}l|}
\hline
\raisebox{-4mm}[0mm][0mm]{$L$} & \multicolumn{8}{c|}{$-c^{(\hat{f}^T_{B^*}/f_{B^*})}_L/4^L$}
& \multicolumn{2}{c|}{\raisebox{-4mm}[0mm][0mm]{$-{}$cont.}} \\
\cline{2-9}
& \multicolumn{2}{c|}{exact} & \multicolumn{2}{c|}{large $\beta_0$} &
\multicolumn{4}{c|}{asymptotic} &\multicolumn{2}{c|}{} \\
\hline
1 & 0&22 & 0&17 & \multicolumn{4}{c|}{} & \multicolumn{2}{c|}{} \\
2 & 1&04 & 2&46 & \multicolumn{4}{c|}{} & \multicolumn{2}{c|}{} \\
3 & \multicolumn{2}{c|}{} &     12&08 &     42&88&     56&240 & 0&0368 \\
4 & \multicolumn{2}{c|}{} &    213&90 &    688&14&    819&160 & 0&0561 \\
5 & \multicolumn{2}{c|}{} &   3020&83 &  13361&95&  15181&120 & 0&1036 \\
6 & \multicolumn{2}{c|}{} &  68980&43 & 310536&81& 343461&096 & 0&2288 \\
\hline
\end{tabular}
\end{table}

If we neglect subleading $\alpha_s$ corrections, then,
from~(\ref{rst}) and~(\ref{mm3}),
\begin{equation}
\hat{f}^T_{B^*}/f_{B^*} = \left(\hat{f}^P_B/f_B\right)^{-1/3}\,.
\label{Dav1}
\end{equation}
This equality also holds at the first order in $1/\beta_0$,
to all orders of $\alpha_s$.
Therefore, the ratio of the perturbative coefficients~(\ref{sAsN}) and~(\ref{mAsN})
is $-\frac{1}{3}$, up to corrections suppressed by $1/n$ and $1/\beta_0$.
Similarly, if we neglect subleading $\alpha_s$ corrections,
including those suppressed by $[\alpha_s/(4\pi)]^{\gamma_{m0}/(2\beta_0)}$,
then, from~(\ref{rff}) and~(\ref{mm3}),
\[
f_{B^*}/f_B = \left(\hat{f}^P_B/f_B\right)^{-\alpha}\,,\quad
\alpha = - \frac{2}{3} \left(1-\frac{\gamma^m_0}{\gamma_{m0}}\right) = \frac{14}{27}\,.
\]
Therefore, the leading asymptotics of the perturbative series
for $f_{B^*}/f_B$~(\ref{fAsN}) and $m/\hat{m}$~(\ref{mAsN}) are related by
\begin{equation}
f_{B^*}/f_B = \left(m/\hat{m}\right)^{-\alpha}\,.
\label{David}
\end{equation}
The term with $N_2$ in~(\ref{fAsN}) violating this relation
is suppressed not only by $(2\beta_0 n)^{-9/25}$,
but also by a small numerical factor $\frac{2}{7}$.
This approximate relation was first noted empirically
at the 2-loop level in~\cite{BG:95},
with the exponent $\alpha=\frac{1}{2}$, which is very close to $14/27$.

Let's summarize our main results.
\begin{enumerate}
\item
The behaviour of the Borel images of perturbative series
near the leading singularity $u=\frac{1}{2}$
for the matching coefficients~(\ref{Su1}), (\ref{Su2}),
and for the ratios $m/\hat{m}$~(\ref{Summ}), $\hat{f}^T_{B^*}/f_{B^*}$~(\ref{Sust}),
and $f_{B^*}/f_B$~(\ref{Suff}) has been found.
The powers of $\frac{1}{2}-u$ are exact;
further corrections are suppressed by positive integer powers of $\frac{1}{2}-u$.
The normalization factors $N_{0,1,2}$ cannot be found within this approach;
they are some unknown numbers of order unity.
Logarithmic branching is a new feature of this problem;
it follows from the fact that the anomalous dimensions matrices
cannot be diagonalized.
\item
Asymptotics of perturbative coefficients $c_L$ at $L\gg1$
for the matching coefficients~(\ref{asy1}), (\ref{asy2}),
and for the same ratios~(\ref{cnmm}), (\ref{cnst}), (\ref{cnff})
have been found.
The powers of $n=L-1$ are exact;
further corrections are suppressed by positive integer powers of $1/n$.
Logarithmic terms follow from the same property of the anomalous dimensions.
\item
The coefficients $B^1_2$ and $B^\V_2$ of the subleading operator $O_2$
for the currents with $\Gamma=1$ and $\rlap/v$ are related by~(\ref{mm1}).
One-loop results for the generic $\Gamma$~(\ref{Vert1})
and for the tensor current~(\ref{B34hat1}) are also new.
\item
The heavy-quark symmetry relations~\cite{Ne:92} for $B$- and $B^*$-meson
matrix elements of subleading operators
get non-trivial radiative corrections~(\ref{R1}), (\ref{Rm}).
\end{enumerate}

\textbf{Note added}. The three-loop anomalous dimension
of the heavy-light quark current $\tilde{\gamma}_2$
has been calculated recently~\cite{CG:03}.
Therefore, the coefficients $c^\Gamma_2$ in~(\ref{series})
are now known, for all $\Gamma$.

\textbf{Acknowledgements}.
We are grateful to A.~Pineda and T.~Lee for useful discussions.
TM acknowledges the support of the DFG Sonderforschungsbereich
SFB/TR9 ``Computational Particle Physics'', FC is supported by DFG
Graduiertenkolleg ``High Energy Physics and Particle Astrophysics'',
and AG acknowledges support by the German
Ministry for Research BMBF, Contract No.\ 05HT1VKB1.

\appendix

\section{UV renormalon ambiguities in the large-$\beta_0$ limit}
\label{UV}

\begin{figure}[ht]
\begin{center}
\begin{picture}(90,19)
\put(45,9.5){\makebox(0,0){\includegraphics{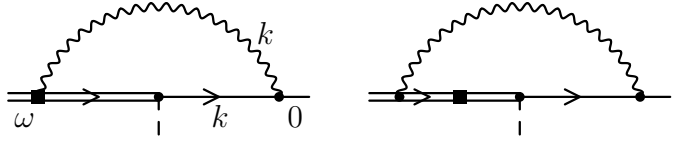}}}
\put(3,2){\makebox(0,0)[b]{$\omega$}}
\put(29,2){\makebox(0,0)[b]{$k$}}
\put(39,2){\makebox(0,0)[b]{$0$}}
\put(35,13){\makebox(0,0)[b]{$k$}}
\end{picture}
\end{center}
\caption{Matrix element of $O_3$;
renormalon chains are inserted into the gluon propagators}
\label{Fig}
\end{figure}

Ultraviolet contributions to the matrix elements of $O_{3,4}$
are independent of the external states,
and we may use quark instead of hadron states (see~(\ref{Hmat1})).
By dimensional analysis, the UV renormalon ambiguities of the matrix elements
of $O_{3,4}$ are proportional to $\Delta\bar{\Lambda}$ times
the matrix element of the lower-dimensional operator $\tilde{\jmath}$
with the same external states.
We consider transition from an off-shell heavy quark
with residual energy $\omega<0$
to a light quark with zero momentum,
this is enough to ensures the absence of IR divergences.
For $O_3$, all loop corrections to the vertex function
(see Fig.~\ref{Fig}) vanish.
The kinetic-energy vertices contain no Dirac matrices,
and we may take $\frac{1}{4}$ of the trace on the light-quark line;
this yields $k^\alpha$ at the vertex,
and the gluon propagator with insertions is transverse.
There is one more contribution~\cite{GN:97}, which we have to take
into account. The matrix element $F$ of $\tilde{\jmath}$
should be multiplied
by the heavy-quark wave-function renormalization $Z_h^{1/2}$,
which contains a kinetic-energy contribution.
This contribution is known to have an UV renormalon ambiguity~\cite{GN:97}
\[
\Delta Z_h = - \frac{3}{2} \frac{\Delta\bar{\Lambda}}{m}\,,
\]
thus giving $-(3/4)(\Delta\bar{\Lambda}/m)F$ as the
ambiguity of the matrix element of $O_3$.
This must be equal to $F\cdot\Delta G_k/(2m)$,
and we recover the first result in~(\ref{DGkm}).

For $O_4$, a straightforward calculation of the diagram
similar to the first one in Fig.~\ref{Fig} gives the bare matrix element
of the usual form (see~\cite{BG:95,GN:97})
\[
\Gamma \left[ 1 + \frac{1}{\beta_0} \sum_{L=1}^\infty
\frac{F(\varepsilon,L\varepsilon)}{L}
\left(\frac{\beta}{\varepsilon+\beta}\right)^L
+ \mathcal{O}\left(\frac{1}{\beta_0^2}\right) \right]
\]
with $\beta=\beta_0\alpha_s(\mu)/(4\pi)$,
\[
F(\varepsilon,u) = - 2 d_\Gamma C_F C_m^{\mathrm{bare}} \frac{\omega}{m}
\left(\frac{\mu}{-2\omega}\right)^{2u} e^{\gamma\varepsilon}
\frac{u\Gamma(-1+2u)\Gamma(1-u)}{\Gamma(2+u-\varepsilon)}
D(\varepsilon)^{\frac{u}{\varepsilon}-1}\,,
\]
where
\begin{eqnarray*}
&&D(\varepsilon) = 6 e^{\gamma\varepsilon}
\Gamma(1+\varepsilon)B(2-\varepsilon,2-\varepsilon) =
1 + \frac{5}{3}\varepsilon+\cdots\,,\\
&&\sigma_{\mu\nu} \Gamma \frac{1+\rlap/v}{2} \sigma^{\mu\nu}
\frac{1+\rlap/v}{2} =
2 d_\Gamma \Gamma \frac{1+\rlap/v}{2}\,,\\
&&d_\Gamma = \frac{1}{2} \left[ \left(1 - 2\eta(n-2)\right)^2 - 3 \right]\,.
\end{eqnarray*}
The renormalization-group invariant matrix element
has the form~(\ref{Laplace}) with $\mu_0=-2\omega e^{-5/6}$ and
\begin{eqnarray*}
&&S(u) = \left.\frac{F(0,u)-F(0,0)}{u}\right|_{\mu=\mu_0}\\
&&{} = - 2 d_\Gamma C_F C_m(-2\omega) \frac{\omega}{m}
\left( \frac{\Gamma(-1+2u)\Gamma(1-u)}{\Gamma(2+u)}
+ \frac{1}{2u} \right)\,.
\end{eqnarray*}
Taking the residue at the pole $u=\frac{1}{2}$,
we find the UV renormalon ambiguity
$(d_\Gamma/3)C_m(-2\omega)(\Delta\bar{\Lambda}/m)$
times the matrix element of $\tilde{\jmath}$.
For $B$-meson, $d_\Gamma=3$;
comparing with~(\ref{fB1}) gives the second result in~(\ref{DGkm}).

\section{The perturbative series for $m/\hat{m}$}
\label{MM}

The ratio of the on-shell mass $m$
and the renormalization-group invariant $\overline{\mathrm{MS}}$ mass $\hat{m}$
is the series
\[
\frac{m}{\hat{m}} = 1 + \sum_{L=1}^\infty c^{(m/\hat{m})}_L
\left(\frac{\alpha_s(\mu_0)}{4\pi}\right)^L
\]
in the $n_l$-flavour $\alpha_s(\mu_0)$ ($\mu_0=e^{-5/6}m$).
Using the relation~\cite{GBGS:90,MR:00}%
\footnote{The three-loop coefficient in it had been found numerically~\cite{CS:99}
before the analytical result~\cite{MR:00} was obtained.}
between $m$ and $\bar{m}(m)$
(omitting the $m_c$ effect known at two loops~\cite{GBGS:90})
together with the 4-loop $\beta$-function~\cite{RVL:97}
and the mass anomalous dimension~\cite{Ch:97,VLR:97},
as well as~\cite{LRV:95,CKS:98}
\[
\alpha_s(m) = \alpha_s'(m)
\left[ 1 + \frac{1}{9} \left( 32 C_A - 39 C_F \right) T_F
\left(\frac{\alpha_s'(m)}{4\pi}\right)^2 + \cdots \right]\,,
\]
we obtain
\begin{eqnarray*}
&&c^{(m/\hat{m})}_1 = C_F \left[ \tfrac{3}{2}
- \left( \tfrac{15}{2} C_F + 8 C_A \right) \frac{1}{\beta_0'}
+ 3 \left( 11 C_F + 7 C_A \right) \frac{C_A}{\beta_0^{\prime2}}
\right]\,,\\
&&c^{(m/\hat{m})}_2 = C_F \Biggl\{ \left( \pi^2 + \tfrac{3}{4} \right) \beta_0'
+ \left( - 6 \zeta_3 - 8 \pi^2 \log 2 + 5 \pi^2 + \tfrac{121}{8} \right) C_F\\
&&\quad{} + \left( 12 \zeta_3 + 4 \pi^2 \log 2 - 5 \pi^2 + \tfrac{149}{16} \right) C_A
+ \left(4\pi^2-\tfrac{46}{3}\right) T_F\\
&&\quad{} + \Bigl[ \tfrac{69}{2} C_F^2
+ \left( 66 \zeta_3 - \tfrac{169}{4} \right) C_F C_A
+ \left( - 66 \zeta_3 + \tfrac{131}{12} \right) C_A^2\\
&&\qquad{} + \tfrac{50}{3} C_F T_F + \tfrac{160}{9} C_A T_F
\Bigr] \frac{1}{\beta_0'}\\
&&\quad{} + \tfrac{1}{48}
\Bigl( 1350 C_F^3 + 504 C_F^2 C_A - 3948 C_F C_A^2 - 483 C_A^3\\
&&\qquad{} - 3520 C_F C_A T_F - 2240 C_A^2 T_F \Bigr)
\frac{1}{\beta_0^{\prime2}}\\
&&\quad{} + \tfrac{3}{2}
\left( - 15 C_F^2 - 5 C_F C_A + 7 C_A^2 \right)
\left( 11 C_F + 7 C_A \right) \frac{C_A}{\beta_0^{\prime3}}\\
&&\quad{} + \tfrac{9}{2} C_F
\left( 11 C_F + 7 C_A \right)^2  \frac{C_A^2}{\beta_0^{\prime4}}
\Biggr\}\,,\\
&&c^{(m/\hat{m})}_3 = C_F \Biggl\{
\left( 12 \zeta_3 + \pi^2 + \tfrac{3}{4} \right) \beta_0^{\prime2}\\
&&\quad{} + \biggl[ \Bigl( 128 a_4 + 100 \zeta_3
+ \tfrac{16}{3} \log^4 2 + \tfrac{32}{3} \pi^2 \log^2 2 - 32 \pi^2 \log 2\\
&&\qquad\quad{} - \tfrac{22}{9} \pi^4 + \tfrac{47}{2} \pi^2 + \tfrac{383}{6} \Bigr) C_F\\
&&\qquad{} + \Bigl( - 64 a_4 - 61 \zeta_3
- \tfrac{8}{3} \log^4 2 - \tfrac{16}{3} \pi^2 \log^2 2 + 16 \pi^2 \log 2\\
&&\qquad\quad{} + \tfrac{2}{9} \pi^4 + \tfrac{13}{3} \pi^2 + \tfrac{1513}{48} \Bigr) C_A\\
&&\qquad{} + \Bigl( 48 \zeta_3 - \tfrac{4}{9} \pi^2 - \tfrac{22}{3} \Bigr) T_F
\biggr] \beta_0'\\
&&\quad{} + \Bigl( 768 a_4 + 80 \zeta_5 + 4 \pi^2 \zeta_3 + 305 \zeta_3
+ 32 \log^4 2 - 32 \pi^2 \log^2 2\\
&&\qquad{} - 508 \pi^2 \log 2
+ \tfrac{4}{3} \pi^4 + \tfrac{673}{3} \pi^2 + \tfrac{7595}{48}
\Bigr) C_F^2\\
&&\quad{} + \Bigl( - 384 a_4 - 200 \zeta_5 + 76 \pi^2 \zeta_3 + 10 \zeta_3
- 16 \log^4 2 + 16 \pi^2 \log^2 2\\
&&\qquad{} + \tfrac{794}{3} \pi^2 \log 2
- \tfrac{2}{3} \pi^4 - \tfrac{1315}{6} \pi^2 - \tfrac{73579}{288}
\Bigr) C_F C_A\\
&&\quad{} + \Bigl( 30 \zeta_5 - 51 \pi^2 \zeta_3 - 150 \zeta_3
- \tfrac{16}{3} \pi^2 \log 2 + \tfrac{5}{2} \pi^4 + \tfrac{139}{6} \pi^2 + \tfrac{431}{8}
\Bigr) C_A^2\\
&&\quad{} + \Bigl( \tfrac{944}{3} \zeta_3 + \tfrac{64}{3} \pi^2 \log^2 2 - \tfrac{896}{9} \pi^2 \log 2
- \tfrac{56}{9} \pi^4 + \tfrac{2602}{27} \pi^2 - \tfrac{2029}{18} \Bigr) C_F T_F\\
&&\quad{} + \Bigl( 40 \zeta_5 - 8 \pi^2 \zeta_3 - \tfrac{424}{3} \zeta_3
- \tfrac{32}{3} \pi^2 \log^2 2 - \tfrac{1856}{9} \pi^2 \log 2\\
&&\qquad{} + \tfrac{28}{9} \pi^4 + \tfrac{4972}{27} \pi^2 - \tfrac{3733}{18}
\Bigr) C_A T_F\\
&&\quad{} + \left( - \tfrac{64}{15} \pi^2 + \tfrac{1640}{27} \right) T_F^2
+ 4 (6\zeta_3+7) C_{FF}\\
&&\quad{} + \biggl[ \Bigl(
- 67 \zeta_3 + 60 \pi^2 \log 2 - \tfrac{75}{2} \pi^2 - \tfrac{7589}{48}
\Bigr) C_F^3\\
&&\qquad{} + \Bigl(
- 440 \zeta_5 + 111 \zeta_3 + 34 \pi^2 \log 2 + \tfrac{61}{2} \pi^2 + \tfrac{14335}{96}
\Bigr) C_F^2 C_A\\
&&\qquad{} + \Bigl(
220 \zeta_5 - 201 \zeta_3 - 32 \pi^2 \log 2 + 61 \pi^2 - \tfrac{4175}{72}
\Bigr) C_F C_A^2\\
&&\qquad{} + \Bigl( 220 \zeta_5 + 175 \zeta_3 + \tfrac{27583}{432} \Bigr) C_A^3\\
&&\qquad{} + \left(-30\pi^2-\tfrac{275}{6}\right) C_F^2 T_F
+ \Bigl( - \tfrac{880}{3} \zeta_3 - 32 \pi^2 + \tfrac{3712}{9} \Bigr) C_F C_A T_F\\
&&\qquad{} + \left(\tfrac{880}{3}\zeta_3+\tfrac{1858}{27}\right) C_A^2 T_F
- \tfrac{1000}{27} C_F T_F^2 - \tfrac{3200}{81} C_A T_F^2\\
&&\qquad{} + 264 (\zeta_3-1) C_{FF} C_A
- 16 (21\zeta_3-2) C_{FA}
\biggr] \frac{1}{\beta_0'}\\
&&\quad{} + \biggl[ - \tfrac{4815}{16} C_F^4
+ \Bigl( - 693 \zeta_3 - 264 \pi^2 \log 2 + 165 \pi^2 + \tfrac{2955}{4} \Bigr) C_F^3 C_A\\
&&\qquad{} + \Bigl( 1205 \zeta_3 - 36 \pi^2 \log 2 - 60 \pi^2
+ \tfrac{61057}{48} \Bigr) C_F^2 C_A^2\\
&&\qquad{} + \Bigl( 428 \zeta_3 + 84 \pi^2 \log 2 - 105 \pi^2
- \tfrac{49189}{96} \Bigr) C_F C_A^3
+ \Bigl( - 616 \zeta_3 + \tfrac{763}{8} \Bigr) C_A^4\\
&&\qquad{} - 125 C_F^3 T_F + \left(132\pi^2-\tfrac{1559}{3}\right) C_F^2 C_A T_F
+ \Bigl( 84 \pi^2 - \tfrac{3775}{9} \Bigr) C_F C_A^2 T_F\\
&&\qquad{} - \tfrac{4739}{18} C_A^3 T_F
+ \tfrac{4400}{27} C_F C_A T_F^2 + \tfrac{2800}{27} C_A^2 T_F^2
+ \Bigl( - \tfrac{3872}{3} \zeta_3 + \tfrac{5324}{9} \Bigr) C_{FF} C_A^2\\
&&\qquad{} + \Bigl( \tfrac{4576}{3} \zeta_3 - \tfrac{1408}{9} \Bigr) C_{FA} C_A
+ \Bigl( - \tfrac{704}{3} \zeta_3 + \tfrac{80}{9} \Bigr) C_{AA}
\biggr] \frac{1}{\beta_0^{\prime2}}\\
&&\quad{} + \biggl[
- \tfrac{1125}{16} C_F^5
+ 1656 C_F^4 C_A
+ \Bigl( 2178 \zeta_3 + \tfrac{785}{8} \Bigr) C_F^3 C_A^2\\
&&\qquad{} - \Bigl( 792 \zeta_3 + \tfrac{132323}{96} \Bigr) C_F^2 C_A^3
- \Bigl( 1386 \zeta_3 + \tfrac{2219}{4} \Bigr) C_F C_A^4
+ \tfrac{147}{4} C_A^5\\
&&\qquad{} + 1100 C_F^3 C_A T_F + \tfrac{3200}{3} C_F^2 C_A^2 T_F
- 280 C_F C_A^3 T_F - \tfrac{980}{3} C_A^4 T_F
\biggr] \frac{1}{\beta_0^{\prime3}}\\
&&\quad{} + \tfrac{1}{16}
\Bigl( 1350 C_F^4 - 2664 C_F^3 C_A - 6140 C_F^2 C_A^2 + 637 C_F C_A^3 + 784 C_A^4\\
&&\qquad{} - 3520 C_F^2 C_A T_F - 2240 C_F C_A^2 T_F \Bigr)
\left( 11 C_F + 7 C_A \right) \frac{C_A}{\beta_0^{\prime4}}\\
&&\quad{} + \tfrac{9}{4} C_F
\left( - 15 C_F^2 + 6 C_F C_A + 14 C_A^2 \right)
\left( 11 C_F + 7 C_A \right)^2 \frac{C_A^2}{\beta_0^{\prime5}}\\
&&\quad{} + \tfrac{9}{2} C_F^2
\left( 11 C_F + 7 C_A \right)^3 \frac{C_A^3}{\beta_0^{\prime6}}\,,
\end{eqnarray*}
where $a_4=\mathop{\mathrm{Li}}\nolimits_4\left(\frac{1}{2}\right)$, and
\[
C_{FF} = \frac{d_F^{abcd}d_F^{abcd}}{T_F^2 N_A}\,,\quad
C_{FA} = \frac{d_F^{abcd}d_A^{abcd}}{T_F N_A}\,,\quad
C_{AA} = \frac{d_A^{abcd}d_A^{abcd}}{N_A}
\]
(see notations in~\cite{RVL:97,VLR:97}).
For $SU(N_c)$ with $T_F=\frac{1}{2}$,
\[
C_{FF} = \frac{N_c^4 - 6 N_c^2 + 18}{24 N_c^2}\,,\quad
C_{FA} = \frac{N_c (N_c^2+6)}{24}\,,\quad
C_{AA} = \frac{N_c^2 (N_c^2+36)}{24}\,.
\]

\section{Renormalization of dimension-4 operators}
\label{R6}

As discussed in~\cite{AN:98}, three operators
\begin{eqnarray}
&&O^\Gamma_1 = \bar{q} i D^\alpha \Gamma h_v\,,\quad
O^\Gamma_2 = \bar{q} \left(-i \overleftarrow{D}_\bot^\alpha\right) \Gamma h_v\,,
\nonumber\\
&&O^\Gamma_3 =
\bar{q} \left(-i v\cdot\overleftarrow{D}\right) \gamma_\bot^\alpha \rlap/v \Gamma h_v
= -i v\cdot\partial \left(\bar{q} \gamma_\bot^\alpha \rlap/v \Gamma h_v\right)
\label{OGamma}
\end{eqnarray}
are closed under renormalization for any $\Gamma$
(note that $O^\Gamma_1-O^\Gamma_2=i\partial_\bot^\alpha\left(\bar{q}\Gamma h_v\right)$).
We have $O^\Gamma_0=Z(\alpha_s(\mu))O^\Gamma(\mu)$, where
\begin{eqnarray}
&&Z = \tilde{Z} + \left(
\begin{array}{ccc}
0 & Z_a & Z_b\\
0 & Z_a & Z_b\\
0 & 0   & 0
\end{array}
\right)\,,
\quad
Z^{-1} = \tilde{Z}^{-1} + \left(
\begin{array}{ccc}
0 & \bar{Z}_a & \bar{Z}_b\\
0 & \bar{Z}_a & \bar{Z}_b\\
0 & 0         & 0
\end{array}
\right)\,,
\nonumber\\
&&\gamma = \tilde{\gamma} + \left(
\begin{array}{ccc}
0 & \gamma_a & \gamma_b\\
0 & \gamma_a & \gamma_b\\
0 & 0        & 0
\end{array}
\right)\,.
\label{structa}
\end{eqnarray}
If we take $\Gamma=\gamma_\alpha \Gamma'$, then
\begin{equation}
O^\Gamma_{10} = O'_{10}\,,\quad
O^\Gamma_{20} = O'_{20}\,,\quad
O^\Gamma_{30} = (3-2\varepsilon) O'_{20}\,,
\label{OO0}
\end{equation}
where
\begin{equation}
O'_1 = i \partial_\alpha \left(\bar{q} \gamma^\alpha \Gamma' h_v\right)\,,\quad
O'_2 = i v\cdot\partial \left(\bar{q} \rlap/v \Gamma' h_v\right)\,.
\label{Op}
\end{equation}
Of course, $O'_0=\tilde{Z}(\alpha_s(\mu))O'(\mu)$.
We obtain
\begin{eqnarray}
&&O^\Gamma_1(\mu) = O'_1(\mu) +
\tilde{Z} \left(\bar{Z}_a+(3-2\varepsilon)\bar{Z}_b\right) O'_2(\mu)\,,
\nonumber\\
&&O^\Gamma_2(\mu) = \left[1 +
\tilde{Z} \left(\bar{Z}_a+(3-2\varepsilon)\bar{Z}_b\right) \right] O'_2(\mu)\,,
\nonumber\\
&&O^\Gamma_3(\mu) = 3 O'_2(\mu)\,.
\label{OO}
\end{eqnarray}
Therefore, $\tilde{Z} \left(\bar{Z}_a+(3-2\varepsilon)\bar{Z}_b\right)$
must be finite at $\varepsilon\to0$.
This allows one to reconstruct $\gamma_b$ from $\gamma_a$:
\begin{equation}
\gamma_b = - \frac{1}{3} \left(\gamma_a + \Delta \gamma_a\right)\,,\quad
\Delta \gamma_a = \frac{1}{3} \gamma_{a0} \left(\gamma_{a0}-2\beta_0\right)
\left(\frac{\alpha_s}{4\pi}\right)^2 + \mathcal{O}(\alpha_s^3)\,.
\label{gammab}
\end{equation}
The anomalous dimensions $\gamma_{a,b}$ has been calculated in~\cite{AN:98}
with the two-loop accuracy (they are called $\gamma_{2,4}$ in this paper):
\begin{eqnarray}
&&\gamma_a = 3 C_F \frac{\alpha_s}{4\pi}
+ C_F \Biggl[ \left(\frac{4}{3}\pi^2-5\right) C_F
\nonumber\\
&&\qquad{} + \left(-\frac{1}{3}\pi^2+\frac{41}{3}\right) C_A
- \frac{10}{3} T_F n_l \Biggr]
\left(\frac{\alpha_s}{4\pi}\right)^2 + \cdots
\label{gammaa}
\end{eqnarray}
The finite parts, at the next-to-leading order, are
\begin{eqnarray}
&&O^\Gamma_1(\mu) = O'_1(\mu)
+ \frac{1}{3} \gamma_{a0} \frac{\alpha_s(\mu)}{4\pi} O'_2(\mu)\,,
\nonumber\\
&&O^\Gamma_2(\mu) = \left[1
+ \frac{1}{3} \gamma_{a0} \frac{\alpha_s(\mu)}{4\pi} \right] O'_2(\mu)\,,
\nonumber\\
&&O^\Gamma_3(\mu) = 3 O'_2(\mu)\,.
\label{OOr}
\end{eqnarray}
Note that $O^\Gamma_{10}=O'_{10}$, but $O^\Gamma_1(\mu)\ne O'_1(\mu)$:
additional counterterms in~(\ref{OO}) yield a finite contribution,
because of the $\mathcal{O}(\varepsilon)$ term in~(\ref{OO0}).

As discussed in~\cite{BNP:01}, two operators
\[
O^k_1 = \bar{q} \left(-iv\cdot\overleftarrow{D}\right) \Gamma h_v\,,\quad
O^k_2 = i \int dx\,T\left\{\bar{q}\Gamma h_v,O_k(x)\right\}
\]
are closed under renormalization for any $\Gamma$, and have
\[
Z = \tilde{Z} +
\left(\begin{array}{cc}0&0\\Z^k&0\end{array}\right)\,,\quad
Z^{-1} = \tilde{Z}^{-1} +
\left(\begin{array}{cc}0&0\\\bar{Z}^k&0\end{array}\right)\,,\quad
\gamma = \tilde{\gamma} +
\left(\begin{array}{cc}0&0\\\gamma^k&0\end{array}\right)\,.
\]
For $\Gamma=1$, these operators are $O_{2,3}$~(\ref{O4});
for $\Gamma=\gamma_\bot^\alpha$, they are $O^\alpha_{4,5}$~(\ref{O6}).
This explains the relevant pieces of~(\ref{gamma4}), (\ref{gamma6}).

Three operators
\begin{eqnarray}
&&O^m_1 = - \tfrac{1}{4} \bar{q} \left(-iv\cdot\overleftarrow{D}\right)
\sigma_{\mu\nu} \Gamma (1+\rlap/v) \sigma^{\mu\nu} h_v\,,
\nonumber\\
&&O^m_2 = - \tfrac{1}{4} \bar{q} \left(-i\overleftarrow{D}_\nu\right)
i\gamma_\mu \rlap/v \Gamma (1+\rlap/v) \sigma^{\mu\nu} h_v\,,
\nonumber\\
&&O^m_3 = i \int dx\,T\left\{\bar{q}\Gamma h_v,O_m(x)\right\}
\label{Om}
\end{eqnarray}
are closed under renormalization for any $\Gamma$.
Note that the indices $\mu$, $\nu$ live in the subspace orthogonal to $v$,
due to $\rlap/v h_v=h_v$.
These operators have~\cite{BNP:01}
\begin{eqnarray}
&&Z = \left(
\begin{array}{ccc}
\tilde{Z} & 0             & 0             \\
Z_b       & \tilde{Z}+Z_a & 0             \\
Z^m_b     & Z^m_a         & \tilde{Z} Z_m
\end{array}
\right)\,,\quad
Z^{-1} = \left(
\begin{array}{ccc}
\tilde{Z}^{-1}  & 0                        & 0                       \\
\bar{Z}_b       & \tilde{Z}^{-1}+\bar{Z}_a & 0                       \\
\bar{Z}^m_b     & \bar{Z}^m_a              & \tilde{Z}^{-1} Z_m^{-1}
\end{array}
\right)\,,
\nonumber\\
&&\gamma = \tilde{\gamma} + \left(
\begin{array}{ccc}
0          & 0          & 0        \\
\gamma_b   & \gamma_a   & 0        \\
\gamma^m_b & \gamma^m_a & \gamma_m
\end{array}
\right)\,.
\label{structm}
\end{eqnarray}
The first two operators in~(\ref{Om}) are~(\ref{OGamma})
$O^{\Gamma'}_3$ and $O^{\Gamma'}_2$ with
$\Gamma'_\alpha=-\frac{1}{4}i\gamma^\mu\rlap/v\Gamma(1+\rlap/v)\sigma_{\mu\alpha}$;
therefore, the upper left $2\times2$ blocks in~(\ref{structm})
follow from~(\ref{structa}).

In the case $\Gamma=1$,
the operators~(\ref{Om}) are related to~(\ref{O4}):
\begin{equation}
O^m_{10} = - (1-\varepsilon) (3-2\varepsilon) O_{20}\,,\quad
O^m_{20} = - (1-\varepsilon) O_{20}\,,\quad
O^m_{30} = O_{40}\,.
\label{Om40}
\end{equation}
We have
\begin{eqnarray}
&&O^m_3(\mu) = O_4(\mu)
\label{Om3a}\\
&&\quad{} + \left[ \tilde{Z}^{-1} Z_m^{-1} Z^m -
(1-\varepsilon) \tilde{Z} \left( \bar{Z}^m_a + (3-2\varepsilon) \bar{Z}^m_b \right)
\right] O_2(\mu)\,.
\nonumber
\end{eqnarray}
Therefore,
\[
\tilde{Z}^{-1} Z_m^{-1} Z^m -
(1-\varepsilon) \tilde{Z} \left( \bar{Z}^m_a + (3-2\varepsilon) \bar{Z}^m_b \right)
\]
must be finite at $\varepsilon\to0$.
This allows one to reconstruct $\gamma^m$ in~(\ref{gamma4})
from $\gamma^m_{a,b}$ in~(\ref{structm}):
\begin{eqnarray}
&&\gamma^m = - \gamma^m_a - 3 \gamma^m_b + \Delta\gamma^m\,,
\label{Dgammam}\\
&&\Delta\gamma^m = \left[
\tfrac{1}{2}\left(\gamma_{m0}-2\beta_0\right)
\left(\gamma^m_{a0}+5\gamma^m_{b0}\right)
- \tfrac{1}{3} \gamma_{a0} \gamma^m_{a0}
\right] \left(\frac{\alpha_s}{4\pi}\right)^2
+ \mathcal{O}(\alpha_s^3)\,.
\nonumber
\end{eqnarray}
The finite part, at the next-to-leading order, is
\begin{equation}
O^m_3(\mu) = O_4(\mu) +
\tfrac{1}{2} \left(\gamma^m_{a0}+5\gamma^m_{b0}\right)
\frac{\alpha_s(\mu)}{4\pi} O_2(\mu)\,.
\label{DOm3a}
\end{equation}

In the case $\Gamma=\gamma_\bot^\alpha$,
the operators~(\ref{Om}) are related to~(\ref{O6}):
\begin{eqnarray}
&&O^m_{10} = (1-\varepsilon) (1+2\varepsilon) O^\alpha_{40}\,,\quad
O^m_{20} = (1-2\varepsilon) O^\alpha_{30} + \varepsilon O^\alpha_{40}\,,
\nonumber\\
&&O^m_{30} = O^\alpha_{60}\,.
\label{Om60}
\end{eqnarray}
We have
\begin{eqnarray}
&&O^m_3(\mu) = O^\alpha_6(\mu)
\label{Om3b}\\
&&\quad{} + \left[ \tilde{Z}^{-1} Z_m^{-1} Z^m_1
+ (1-2\varepsilon) (\tilde{Z}+Z_a) \bar{Z}^m_a \right] O^\alpha_3(\mu)
\nonumber\\
&&\quad{} + \left[ \tilde{Z}^{-1} Z_m^{-1} Z^m_2
+ \left((1-2\varepsilon)Z_b+\varepsilon\tilde{Z}\right) \bar{Z}^m_a
+ (1-\varepsilon) (1+2\varepsilon) \tilde{Z} \bar{Z}^m_b \right] O^\alpha_4(\mu)\,.
\nonumber
\end{eqnarray}
Therefore,
\[
\tilde{Z}^{-1} Z_m^{-1} Z^m_1
+ (1-2\varepsilon) (\tilde{Z}+Z_a) \bar{Z}^m_a
\]
and
\[
\tilde{Z}^{-1} Z_m^{-1} Z^m_2
+ \left((1-2\varepsilon)Z_b+\varepsilon\tilde{Z}\right) \bar{Z}^m_a
+ (1-\varepsilon) (1+2\varepsilon) \tilde{Z} \bar{Z}^m_b
\]
must be finite at $\varepsilon\to0$.
This allows one to reconstruct $\gamma^m_{1,2}$ in~(\ref{gamma6})
from $\gamma^m_{a,b}$ in~(\ref{structm}):
\begin{eqnarray}
&&\gamma^m_1 = \gamma^m_a + \Delta\gamma^m_1\,,\quad
\gamma^m_2 = \gamma^m_b + \Delta\gamma^m_2\,,
\label{Dgammam12}\\
&&\Delta\gamma^m_1 = \gamma^m_{a0}
\left(\gamma_{a0}-\gamma_{m0}+2\beta_0\right)
\left(\frac{\alpha_s}{4\pi}\right)^2
+ \mathcal{O}(\alpha_s^3)\,,
\nonumber\\
&&\Delta\gamma^m_2 = \left[
\tfrac{1}{2} \left(\gamma_{m0}-2\beta_0\right)
\left(\gamma^m_{a0}+\gamma^m_{b0}\right)
- \tfrac{1}{3} \gamma_{a0} \gamma^m_{a0} \right]
\left(\frac{\alpha_s}{4\pi}\right)^2
+ \mathcal{O}(\alpha_s^3)\,.
\nonumber
\end{eqnarray}
The finite part, at the next-to-leading order, is
\begin{equation}
O^m_3(\mu) = O^\alpha_6(\mu)
- \gamma^m_{a0} \frac{\alpha_s(\mu)}{4\pi} O^\alpha_3(\mu)
+ \tfrac{1}{2} \left(\gamma^m_{a0}+\gamma^m_{b0}\right)
\frac{\alpha_s(\mu)}{4\pi} O^\alpha_4(\mu)\,.
\label{DOm3b}
\end{equation}

The anomalous dimensions $\gamma^m_{1,2}$
has been calculated in~\cite{BNP:01} with the two-loop accuracy
(they are called $\gamma^{\mathrm{mag}}_{3,1}$ in this paper):
\begin{eqnarray}
&&\gamma^m_1 = -2 C_F \frac{\alpha_s}{4\pi}
+ C_F \Biggl[ \left(-\frac{8}{9}\pi^2+\frac{4}{3}\right) C_F
\nonumber\\
&&\qquad{} + \left(\frac{2}{9}\pi^2-\frac{206}{9}\right) C_A
+ \frac{52}{9} T_F n_l \Biggr]
\left(\frac{\alpha_s}{4\pi}\right)^2 + \cdots
\nonumber\\
&&\gamma^m_2 = -2 C_F \frac{\alpha_s}{4\pi}
+ C_F \Biggl[ \left(-\frac{40}{9}\pi^2-\frac{10}{3}\right) C_F
\nonumber\\
&&\qquad{} + \left(\frac{10}{9}\pi^2+\frac{46}{9}\right) C_A
- \frac{44}{9} T_F n_l \Biggr]
\left(\frac{\alpha_s}{4\pi}\right)^2 + \cdots
\label{Mix2}
\end{eqnarray}
The anomalous dimension $\gamma^m$~(\ref{Mix}) is,
from~(\ref{Dgammam}) and~(\ref{Dgammam12}),
\begin{eqnarray}
&&\gamma^m = - \gamma^m_1 - 3 \gamma^m_2 + \Delta\gamma\,,
\label{Dgammamm}\\
&&\Delta\gamma = \left[
\left(\gamma_{m0}-2\beta_0\right)
\left(\gamma^m_{10}+4\gamma^m_{20}\right)
- \tfrac{1}{3} \gamma_{a0} \gamma^m_{10}
\right] \left(\frac{\alpha_s}{4\pi}\right)^2
+ \mathcal{O}(\alpha_s^3)\,.
\nonumber
\end{eqnarray}

\end{document}